\documentclass{aa}  

\usepackage{graphicx}
\usepackage{txfonts}
\usepackage{color}
\usepackage{listings}
\usepackage{array}
\usepackage{subfigure}
\usepackage{hyperref}
\usepackage{comment}
\usepackage[capitalize,nameinlink]{cleveref}
\usepackage{threeparttable}

\newcommand\inputpgf[2]{{
\let\includegraphicsWithoutPath\includegraphics
\renewcommand{\includegraphics}[2][]{\includegraphicsWithoutPath[##1]{#1/##2}}
\input{#1/#2}
}}

\DeclareUnicodeCharacter{2212}{-}

\begin{document}

\title{Migration of pairs of giant planets in low-viscosity discs.}

\author{P. Griveaud \inst{1}
\and A. Crida \inst{1} 
\and E. Lega\inst{1}}

\institute{Universit{\'e} C{\^o}te d'Azur, Observatoire de la C{\^o}te d'Azur, CNRS, Laboratoire Lagrange, France}

        \abstract
        {When considering the migration of Jupiter and Saturn, a classical result is to find the planets migrating outwards and locked in the 3:2 mean motion resonance (MMR). These results were obtained in the framework of viscously accreting discs, in which the observed stellar accretion rates constrained the viscosity values. However, it has recently been shown observationally and theoretically that discs are probably less viscous than previously thought.}
        {Therefore, in this paper, we explore the dynamics of pairs of giant planets in low-viscosity discs.}
        {We performed two-dimensional hydrodynamical simulations using the grid-based code FARGOCA.}
        {In contrast to classical viscous discs, we find that the outer planet never crosses the 2:1 resonance and the pair does not migrate outwards. After a wide parameter exploration, including the mass of the outer planet, we find that the planets  are primarily locked in the 2:1 MMR and in some cases in the 5:2 MMR. We explain semi-analytically why it is not possible for the outer planet to cross the 2:1 MMR in a low-viscosity disc.} 
        {We find that pairs of giant planets migrate inwards in low-viscosity discs. Although, in some cases, having a pair of giant planets can slow down the migration speed with respect to a single planet. Such pairs of slowly migrating planets may be located, at the end of the disc phase, in the {population of exoplanets of 'warm Jupiters'}. However, the planets never migrate outwards. These results could have strong implications on the Solar System's formation scenarios if the Sun's protoplanetary disc had a low viscosity.}

       \maketitle

\section{Introduction}\label{sec:intro}

{The migration history of Jupiter and Saturn played a significant role in the shaping of our Solar System \citep{walsh_low_2011, tsiganis_origin_2005}. To date, there are 163 giant exoplanets with masses between Saturn and five Jupiter masses and an orbital period larger than 100 days. Within this population of so-called warm Jupiters, there are 43 systems composed of at least two giant planets.\footnote{From the \href{https://github.com/OpenExoplanetCatalogue/open_exoplanet_catalogue}{Open Exoplanet Catalogue} by \cite{Rein2012}.} Studying the dynamical history of such systems during their formation is a key point in understanding the variety of architectures of planetary systems \citep[see][for a review]{zhu_exoplanet_2021}.} \\
{In the specific case of our Solar System, \cite{masset_reversing_2001} found} that the progenitors of Jupiter and Saturn may have undergone convergent migration until Saturn was captured in the 3:2 mean motion resonance (MMR) with Jupiter. In such a compact orbital configuration, the two planets are observed to revert their direction of migration and start migrating outwards while remaining locked in the 3:2 resonance. The work of Masset and Snellgrove was extended by \cite{morbidelli_dynamics_2007-1}, who explored a wider set of initial conditions and disc parameters. \\
Focussing on the specific case of our Solar System, \cite{walsh_low_2011} suggested that the Jupiter-Saturn pair did reverse their migration and that this phase of outward migration started when Jupiter was at a distance of about $1.5$ {au} from the Sun. This scenario, labelled {Grand Tack}, would explain both the small mass of Mars with respect to the Earth and the variation of composition in the asteroid belt, {which are key points in explaining the final architecture of our Solar System}. \\
{The reversal of migration, commonly called the Masset and Snellgrove mechanism, requires the following:}
\begin{itemize}
    \item[\textit{(i)}] The mass of the interior planet must exceed that of the exterior one; 
    \item[\textit{(ii)}] The two planets must share a common gap so that the outer (negative) Lindblad torque exerted on the less massive planet does not overcome the (positive) inner Lindblad torque exerted on the more massive one;
    \item[\textit{(iii)}] The orbital configuration of the pair of planets must be compact such that their horseshoe regions are close enough to overlap. In this way the gas that enters the horseshoe region outside the outer planet's orbit may be transported in the disc region inside the orbit of the inner planet, passing through its horseshoe region. The pileup of gas at the inner edge of the gap then maintains the positive inner Lindblad torque that sustains the outward migration phase.
\end{itemize}
{These three conditions are often met when considering a Jupiter-Saturn pair locked in 3:2 MMR. Furthermore,} \cite{crida_long_2009} showed that the outward migration can then be maintained on a very long range.\\
\cite{dangelo_outward_2012} also addressed {the question of the migration of Jupiter and Saturn in a more complex thermodynamical model, where the gas disc is evolving and planets are accreting gas. They showed }that there can be severe limitations to the outward migration conditions: the relative speed of the pair of planets may not be large enough for the exterior planet to cross the 2:1 resonance{. In such a configuration, the conditions \textit{(ii)} and \textit{(iii)} are not satisfied.} The authors {also} showed that the specific requirement on the planets' masses {(i.e. condition \textit{(i)})} may easily be violated by planetary accretion. More recently, \cite{chametla_capture_2020} have suggested that the conditions for the Grand Tack model depend on the precise initial orbit of Jupiter and Saturn and on their formation {timescales}. They find that capture in the 2:1 MMR and slow inward migration is favoured{ if Saturn enters in resonance with Jupiter before being fully grown, at about two-thirds of its final mass.} \\
{At odds with these results,} \cite{pierens_outward_2014} found that in relatively small mass, viscously heated discs {(similar model to \cite{dangelo_outward_2012})}, Jupiter and Saturn may migrate outwards when locked in the 2:1 MMR. {However,} this result was obtained {for viscosity values lower than the one used in \cite{dangelo_outward_2012}. This implies} that discs are characterised by an {unusually small} aspect ratio of the order of 0.02-0.03. As the authors noticed, discs are not expected to be so cold when adding stellar irradiation {(although cold stellar irradiated discs have been recently discussed in \cite{savvidou_growth_2021})}, so that outward migration in the 2:1 MMR cannot be maintained on a very long range. \\
{At the time of these cited studies,} discs were believed to have a non-negligible viscosity to justify the high accretion rates of gas onto the central star. Viscosity was raised by the turbulence generated via the magneto rotational instability {(MRI)} \citep{balbus_powerful_1991}. Therefore, {the previously cited} literature concerning the migration of a pair of giant planets mainly {focusses} on discs with a value for the $\alpha$ turbulence parameter \citep{shakura_black_1973} in the interval $[10^{-3}, 10^{-2}]$. \\
However, it has recently been shown, {  both} theoretically and observationally, that protoplanetary discs are probably less viscous than previously thought.
{On the theoretical side, } for example, it has been shown that the gas near the mid-plane has a level of ionisation that is too weak to sustain the MRI \citep{gammie_layered_1996}. {And more} recently, the inclusion of non-ideal {magnetohydrodynamics (MHD)} effects, such as ambipolar diffusion, led to the conclusion that disc should not be turbulent even at {the} surface (see \cite{turner_transport_2014} for a review). 
 {On the observational side, }
\cite{pinte_dust_2016} derived a value of the turbulence parameter on HL Tau of the order of $10^{-4}$ by modelling dust settling. With the same method, \cite{villenave_highly_2022} found a value of the order of $10^{-5}$. Another example of turbulent parameter $<10^{-3}$ comes from observations of the Lupus and Upper Sco star-forming regions {\citep{sellek_dusty_2020}, where }such low viscosity values are required in order to reproduce the range of masses and accretion rates seen. \\
In this context, we consider it important to revisit the migration of pairs of giant planets in low-viscosity discs. In this paper, we consider two-dimensional discs. We are aware of the fact that low-viscosity discs do not provide the sufficient transport of gas required for a typical accretion rate observed in young stars \citep{hartmann_accretion_1998,manara_evidence_2016}. Magnetically driven disc winds have been proposed as a mechanism to remove angular momentum from thin ionised surface layers of low-viscosity protoplanetary discs \citep{suzuki_disk_2009, bai_wind-driven_2013,turner_transport_2014,bethune_global_2017}. Angular momentum removal promotes, in these layers, a fast radial transport of gas towards the central star. We will mimic the effect of magnetically driven winds in three-dimensional hydrodynamical simulations (following the model in \cite{lega_migration_2022}) in a forthcoming paper. In this paper, we prefer to explore the dynamics in two-dimensional simulations, which have the advantage of providing a wide parameter exploration over long {timescales} in reasonable computation time.\\
The layout of this paper is as follows. We first present, in \cref{sec:ModelSetUp}, how we conducted our study, including the description of the physical model and the numerical setup, as well as the structure of our simulations. We then present our results in \cref{sec:results}, with each subsection detailing the parameter exploration of this study. In {\cref{subsec:criterion}}, we explain why Saturn never crosses the 2:1 MMR in low-viscosity discs using a semi-analytical method. We confirm the robustness of these results for a wider range of planetary masses in \cref{sec:Mp}. Finally, we discuss our work and conclude in \cref{sec:discussion}.

\section{Model setup}\label{sec:ModelSetUp}

\subsection{Physical model}\label{sec:phys_model}

In this work, we consider a protoplanetary non-self-gravitating disc made of gas only. The gas follows the vertically integrated Navier-Stokes equations \citep{masset_co-orbital_2002}. Similar to \cite{lega_migration_2021}, we used an adiabatic equation of state to which we added an exponential damping of the temperature perturbations. This damping was done on the timescale of the cooling time $\tau_c$. Following {\cite{lega_migration_2021}}, we considered $\tau_c$ to be equal to one local orbital period.  \\
The two-dimensional system is described in polar coordinates $(r,\phi)$, centred on the star, where $r$ is the radial coordinate and $\phi$ the azimuthal coordinate. 
We adopted a flared aspect ratio given by \cite{chiang_spectral_1997}
such that $h=h_0(r/r_0)^{2/7}$ with $h_0=0.05$, unless otherwise stated, {and $r_0$ is the unit length (see \cref{sec:units})}.
The disc density is given by $\Sigma = \Sigma_0 (r/r_0)^{-1/2}$ where $\Sigma_0 = 6.76\cdot 10^{-4}$ in code units for the nominal case. This corresponds to $222 \,\text{g cm}^{-2}$ for $r_0 = 5.2$ {au} and {a Sun mass star.} 
We used the $\alpha$ prescription \citep{shakura_black_1973} for the kinematic viscosity, such that $\nu = \alpha c_s H$, {where $c_s$ is the sound speed and $H$ the scale height of the disc such that $H=hr$}. Throughout this paper we use $\alpha= 10^{-4}$. This value is observationally and theoretically motivated as mentioned in the Introduction. Moreover from a numerical point of view, it has been shown that numerical convergence cannot be obtained for values of $\alpha$ lower than $5\cdot 10^{-5}$ \citep{mcnally_migrating_2019}. Numerical convergence with respect to the viscosity is discussed {in \cref{app:Resolution}}. \\
Two planets were added in the system at a different time. The inner planet is a Jupiter-mass planet with mass $M_J/M_\star = q_J = 10^{-3}$. The outer planet, unless otherwise stated, is a Saturn-mass planet with mass $M_S/M_\star = q_S = 2.9 \cdot 10^{-4}$. For simplicity, the two planets are henceforth referred to as Jupiter and Saturn, respectively. \\

\subsection{Units}\label{sec:units}

The code units are such that $G=M_\star=r_0=1$, where $r_0$ is the unit length and the initial semi-major axis of Jupiter. Therefore, the orbital period at $r_0 = 1$ is $2\pi$ time code units. Planetary masses{, $M_p$,} are then given as a ratio to the stellar mass $q_p = M_p/M_\star$.\\
We remark that the Navier-Stokes equations are invariant with respect to the mass of the system, even in the case of an adiabatic equation of state with our prescribed cooling (see \cref{sec:phys_model}), since the energy is directly proportional to the density. Only the mass ratios matter\,; it would not be the case if the cooling was given by the opacity of the disc, which depends on the amount of dust in physical units. \\
{For example the three following sets of physical parameters correspond to one single simulation:
\begin{itemize}
    \item[\textit{(i)}] $M_\star=M_\odot$, $M_p=M_{ J}$, $\Sigma_0 = 1000$\,g.cm$^{-2}$, $r_0=5$ au
    \item[\textit{(ii)}] $M_\star=2M_\odot$, $M_p=2M_{ J}$, $\Sigma_0 = 2000$\,g.cm$^{-2}$, $r_0=5$ au
    \item[\textit{(iii)}] $M_\star=M_\odot$, $M_p=M_{ J}$, $\Sigma_0 = 250$\,g.cm$^{-2}$, $r_0=2.5$ au
\end{itemize}
Indeed, in the three cases, $M_\star = 1$, $M_p=10^{-3}$, and $r_0=1$ {are in code units and therefore $\Sigma_0 = 3\cdot 10^{-3} M_\star r_0^{-2}$}. In the case of \textit{(ii)}, the mass unit is doubled with respect to case \textit{(i),} while the distance unit is unchanged. In the case of \textit{(iii)}, the mass unit is unchanged while the distance unit is halved, such that the surface density is divided by four.
\footnote{{More detailed calculations: \\ 
if in \textit{(i)}, $\Sigma_0=3\cdot 10^{-3} \left(M_\odot \cdot (5 \textrm{au})^{-2} \right) = 3\cdot 10^{-3} \times 2\cdot 10^{33} \rm g \times 1.78 \cdot 10^{-28} cm^{-2}  \approx 1000$\,g.cm$^{-2}$; 
\\ then in \textit{(ii)}, $\Sigma_0=3\cdot 10^{-3} \left( 2M_\odot \cdot (5 \textrm{au})^{-2} \right) = 3\cdot 10^{-3} \times 4\cdot 10^{33} \rm g \times 1.78 \cdot 10^{-28} cm^{-2}  \approx 2000$\,g.cm$^{-2}$; 
\\ and in \textit{(iii)}, $\Sigma_0=3\cdot 10^{-3} \left( M_\odot \cdot (2.5 \textrm{au})^{-2} \right) = 3\cdot 10^{-3} \times 2\cdot 10^{33} \rm g \times 4.47 \cdot 10^{-29} cm^{-2}  \approx 250$\,g.cm$^{-2}$}.}} \\
In other words, only the dimensionless quantities {$q$}, $\alpha$, and $h_0$ {and the disc-to-star mass ratio $\mu = \pi\Sigma_0{r_0}^{2}/M_\star$ \citep[see also][]{crida_cavity_2007}} are relevant to characterise a simulation. Similarly, the initial position of Saturn matters only in terms of the ratio to the semi-major axis of Jupiter. Thus, these are the parameters given in {Table} \ref{tab:simulations} below.
The reader can then interpret the result choosing their favourite stellar mass and orbital distance\footnote{{Only for a system governed by the equations that we simulated and in which no other physical phenomena have significant effects (e.g. UV photo-evaporation by the central star).}}. To give some sense of the {timelines} of the phenomenons that we observed in the frame of planetary formation, in this paper, we provide numbers in physical units based on $M_\star=M_\odot$ and $r_0=5.2$ {au}, so that one orbital period {at $r_0$} corresponds to $11.85$ years.

\subsection{Numerical setup}\label{sec:num_setup}

The 2D simulations were carried out with the code FARGOCA, the code FARGO \citep{masset_fargo_2000} with Co-latitude Added \citep{lega_migration_2014}\footnote{The simulations presented in this paper have been obtained with a recently re-factorised version of the code that can be found at \url{https://disc.pages.oca.eu/fargOCA/public/}}. The code was parallelised using a hybrid combination of MPI between the nodes and OpenMP on shared memory multi-core processors.\\
The radial domain extends in the range $r\in [0.2,9]r_0$, which corresponds to $r\in [1.04,46.8]$ {au} for $r_0=5.2$ {au}. The mesh was divided in $N_{\text{rad}}=568$ radial cells with logarithmic spacing and $N_{\text{sec}}=940$ azimuthal cells. The resolution is such that $dr/r = d\theta = 0.0067${, so that a pressure scale length $H$ is covered by about five cells at the inner edge of the grid. This resolution is proven to be sufficient in Appendix~\ref{app:Resolution}}. The boundary conditions used in the radial direction follow the prescription of the evanescent boundary condition \citep{de_val-borro_comparative_2006}.\\
We worked in the reference frame co-rotating with Jupiter so that Jupiter laid on the $x$-axis at all times. We recall that the system is centred on the star and, therefore, the star is accelerated by the gravity of the planets and the disc. Indirect forces thus arise and must be carefully taken into account. The planets feel the indirect forces that are due to their own gravity as well as that of the disc. The disc feels indirect forces from the planets. Since the self-gravity of the disc has not been accounted for, we did not account for the indirect forces of the gas onto itself \citep[and in prep]{Crida+2022_SF2A}.
The gravitational potential of the planets on the disc was modelled as follows: 
\begin{equation}
    \psi_p = -\frac{G M_p}{\sqrt{d^2 + \epsilon^2}} \,,
\end{equation}
where $d$ is the distance between the planet and a grid cell centre. The smoothing parameter $\epsilon$ used to evaluate the potential in the vicinity of the planets scales with the scale height of the disc $H$ such that $\epsilon = 0.6H$ as is usually done in 2D simulations (e.g. \cite{crida_dynamical_2009}).

\subsection{Simulation setup}\label{sec:sim_setup}

\begin{figure}[h]
    \centering
    \includegraphics[width=\columnwidth]{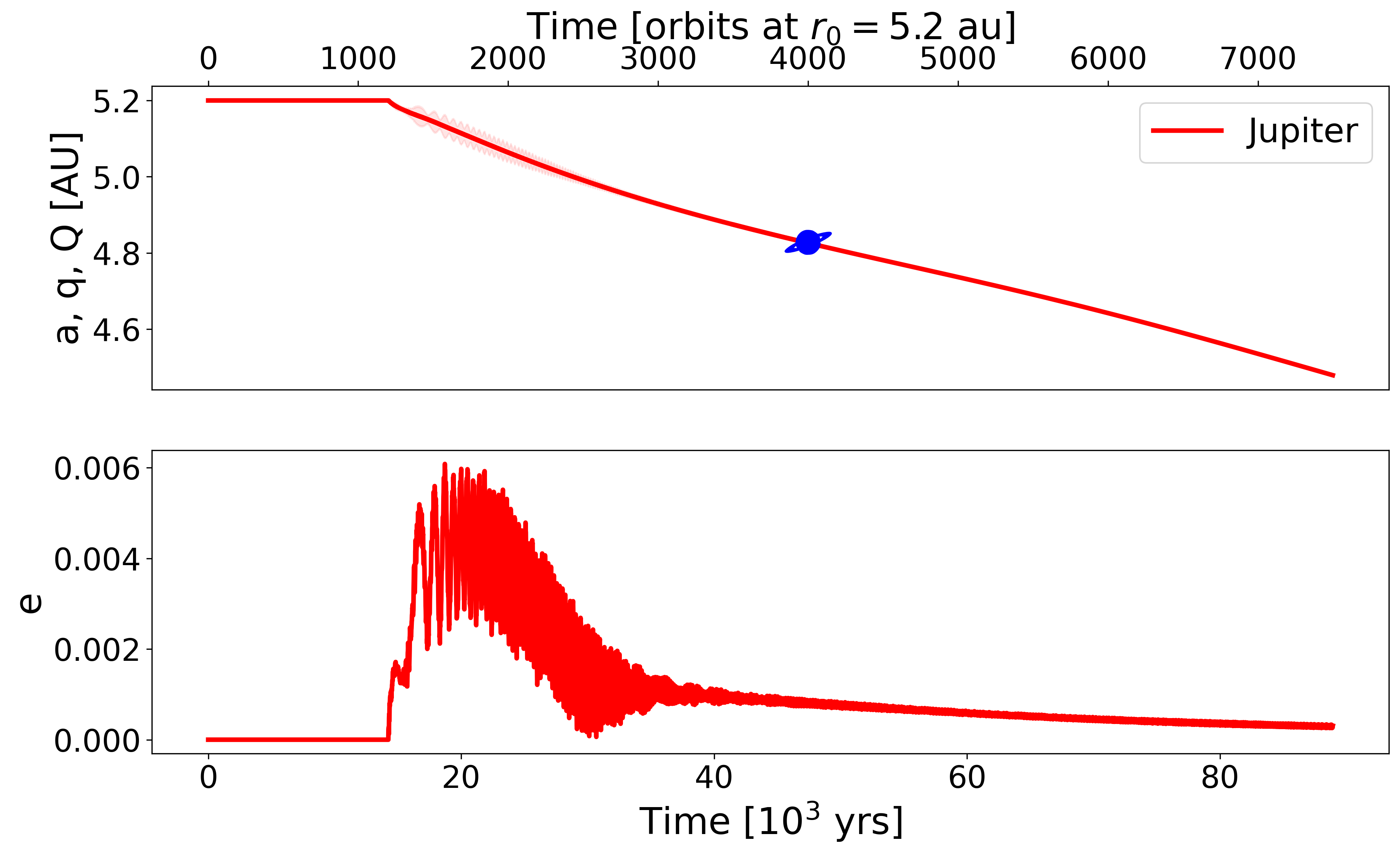}
    \caption{Migration of Jupiter alone in a low-viscosity disc{, corresponding to {the parameters of the nominal simulation $N$}}. The top panel shows the evolution of the semi-major axis of the planet and the bottom panel is its eccentricity. The blue Saturn-shaped marker in the top panel indicates the moment at which Saturn was added in the simulations presented in \cref{sec:results}.}
    \label{fig:Jsing}
\end{figure}

The simulations we present in this paper proceeded in two phases.  During the first phase, {we let Jupiter evolve alone in the disc. We started by introducing the planet on a fixed circular orbit at a distance $a_J = r_0$ from the star, with the remaining orbital elements being initialised at zero. Its mass was increased analytically\footnote{{i.e. the planet does not accrete gas from the disc.}} from zero up to $q_J$ in $800$ local orbits, following}
\begin{align}\label{eq:massTaper}
    q(t) = q_J \sin^2\left(\frac{\pi}{2} \frac{t}{T_{\rm growth}}\right) \,,
\end{align}
{where $T_{\rm growth}=2\pi\times 800 \left(\frac{a_J}{r_0}\right)^{3/2}$ code time units.}
{We waited an additional $400$ orbits to let the disc stabilise before allowing Jupiter to migrate. This process avoids triggering instabilities that could arise from introducing a Jupiter mass planet in the disc directly \citep{hammer_slowly-growing_2017, hallam_constraining_2020}. If Jupiter was allowed to grow and migrate at the same time, it would have a fast inward migration episode before reaching the type II regime. This would shift the positions of the {planets to the inside of the disc} and as explained in \cref{sec:units} the results would only have to be re-scaled.} {The initial phase of the migration of Jupiter alone in the {nominal} disc is shown in \cref{fig:Jsing}.} \\
The second phase started when introducing the outer planet in the disc after $4000$ initial orbital periods of Jupiter, corresponding to about $47\,500$ years. {We call this time $T_{0,S}$. Saturn was also initialised on a circular orbit at a distance $a_S$ determined by the ratio $\left[\frac{a_S}{a_J} \right]_{T_{0,S}}$.} {Adding the outer planet at this moment was motivated by the fact that planet formation is favoured to take place at pressure bumps located at the edge of planetary gaps \citep[and references therein]{eriksson_fate_2021}. We then increased the outer planet mass from zero to Saturn's mass, with the same method as in \cref{eq:massTaper}} for 800 initial orbital periods {of Saturn} (corresponding to about $25\,200$ years in our nominal case, $\left[\frac{a_S}{a_J} \right]_{T_{0,S}}=2$). \\
The planet was free to migrate during its growth. This means that the outer planet { goes through} a phase of type I migration before {it reaches} a mass significant enough to open a gap. This allowed the planet to rapidly migrate and possibly be caught in resonance with Jupiter. The case of Saturn growing on a fixed orbit and then being allowed to migrate has been studied, but it resulted in both planets opening separate gaps which do not overlap. The planets' evolution is independent with very slow migrations. 

\section{Simulation results} \label{sec:results}

In the following section, we present the results of the migration of the pair of giant planets for different disc parameters and separations. {Table} \ref{tab:simulations} gathers the simulations that were run for this study, the parameters that were explored, and a summary of their results. 

\begin{table}
\begin{threeparttable} 
    \caption{Summary of the simulation parameters explored in this study. Column 1 - Simulation name. Column 2 - Parameters: mass and starting position of the outer planet {with respect to the position of the inner planet}, disc aspect ratio, and disc surface density. Column 3 - Mean motion resonance (MMR) of the planets at the end of the simulation. Bold parameters indicate the changed values with respect to the nominal case $N$.}\label{tab:simulations}
    \centering
    \begin{tabular}{c c c c c c}
    \hline\hline
     {Name} & \multicolumn{4}{c}{{Parameters}} & \multicolumn{1}{c}{{Results}} \\ %
      & \raisebox{-1.5ex}{$M_p$} 
      & \raisebox{-1.5ex}{\large{$\left[\frac{a_S}{a_J} \right]_{T_{0,S}}$}} 
      & \raisebox{-0.5ex}{$h_0$} 
      & \raisebox{-0.5ex}{$\Sigma_0$} 
      & \raisebox{-1.5ex}{MMR} \\[-1.5ex]
      & 
      &  
      & \multicolumn{2}{c}{(code units)} 
      & \\[0.5ex] 
      \hline
    $N$               & $M_S$               &    2          & 0.05           & 6.76e-4          & {2:1}    \\
    $C$               & $M_S$               &    2          & \textbf{0.035} & 6.76e-4          & {5:2}    \\
    $H$               & $M_S$               &    2          & \textbf{0.06}  & 6.76e-4          & {2:1}    \\
    {$N_{\Sigma -}$}  & $M_S$               &    2          & 0.05           & \textbf{1.61e-4} & {2:1}    \\
    {$N_{\Sigma +}$}  & $M_S$               &    2          & 0.05           & \textbf{2.8e-3}  & {2:1}    \\
    {$C_{\Sigma -}$}  & $M_S$               &    2          & \textbf{0.035} & \textbf{1.61e-4} & 2:1      \\
    {$C_{\Sigma +}$}  & $M_S$               &    2          & \textbf{0.035} & \textbf{2.8e-3}  & 2:1      \\
    {$C_{2.5}$}       & $M_S$               & \textbf{2.5}  & \textbf{0.035} & 6.76e-4          & {5:2}    \\
    {$C_3$}           & $M_S$               & \textbf{3}    & \textbf{0.035} & 6.76e-4          & 2:1      \\
    $N_{M_J}$         & $\boldsymbol{M_J}$  &    2          & 0.05           & 6.76e-4          & 2:1      \\main
    $N_{2M_J}$        & $\boldsymbol{2M_J}$ &    2          & 0.05           & 6.76e-4          & 2:1\tnote{*} \\
    $N_{3M_J}$        & $\boldsymbol{3M_J}$ &    2          & 0.05           & 6.76e-4          & 2:1      \\ \hline
    \end{tabular}
    \begin{tablenotes}
        \item[*] Unstable configuration {after 130} kyr.
    \end{tablenotes}
    \end{threeparttable}
\end{table}

        \subsection{Nominal simulation}

\begin{figure*}[t]
    \centering
    \includegraphics[width=\textwidth]{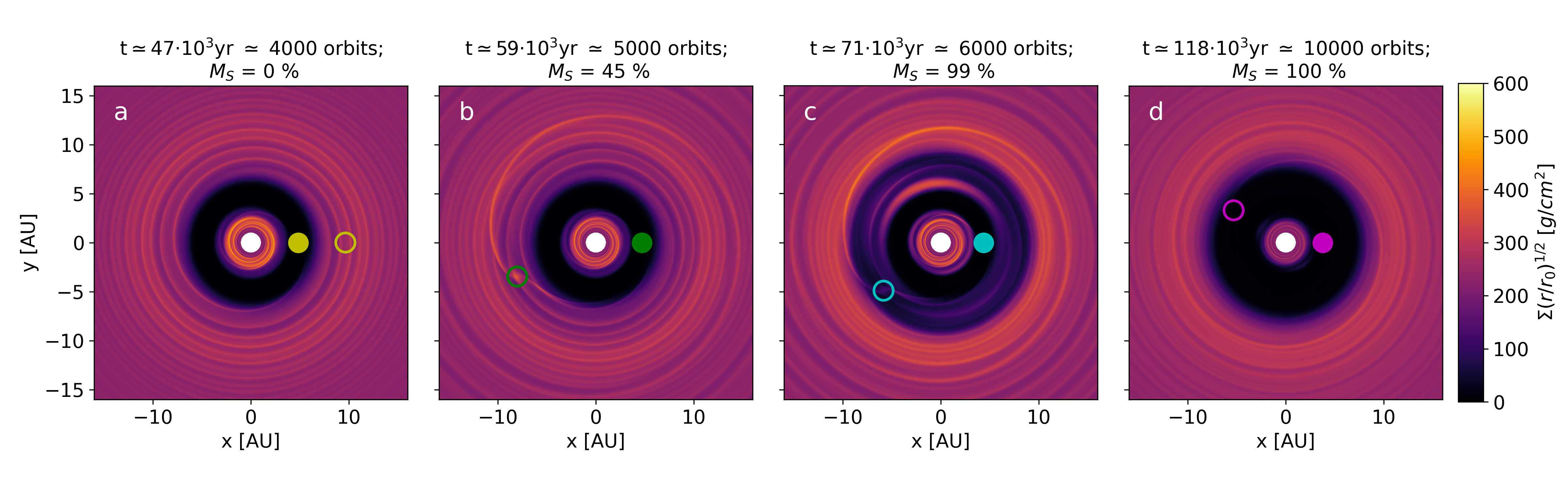}
    \caption{Surface density of the disc at different {times} of the nominal simulation. The filled and empty circles mark the positions of Jupiter and Saturn, respectively. Panel a shows the disc at the moment of introduction of Saturn, $T_{0,S}$, while the outer planet still has zero mass. In panel b, the planet is at $45\%$ of its total mass. In panel c, the planet is just about to reach  its final mass. Panel d shows the disc and the planets {closer to} the end of the simulation. At this point both planets are in a large common gap and locked in the {2:1 MMR}. The colours given to the planets correspond to the markers in \cref{fig:nom_OrbPar}.}
    \label{fig:nom_SigMap}
\end{figure*}

\begin{figure}
    \centering
    \includegraphics[width=\columnwidth]{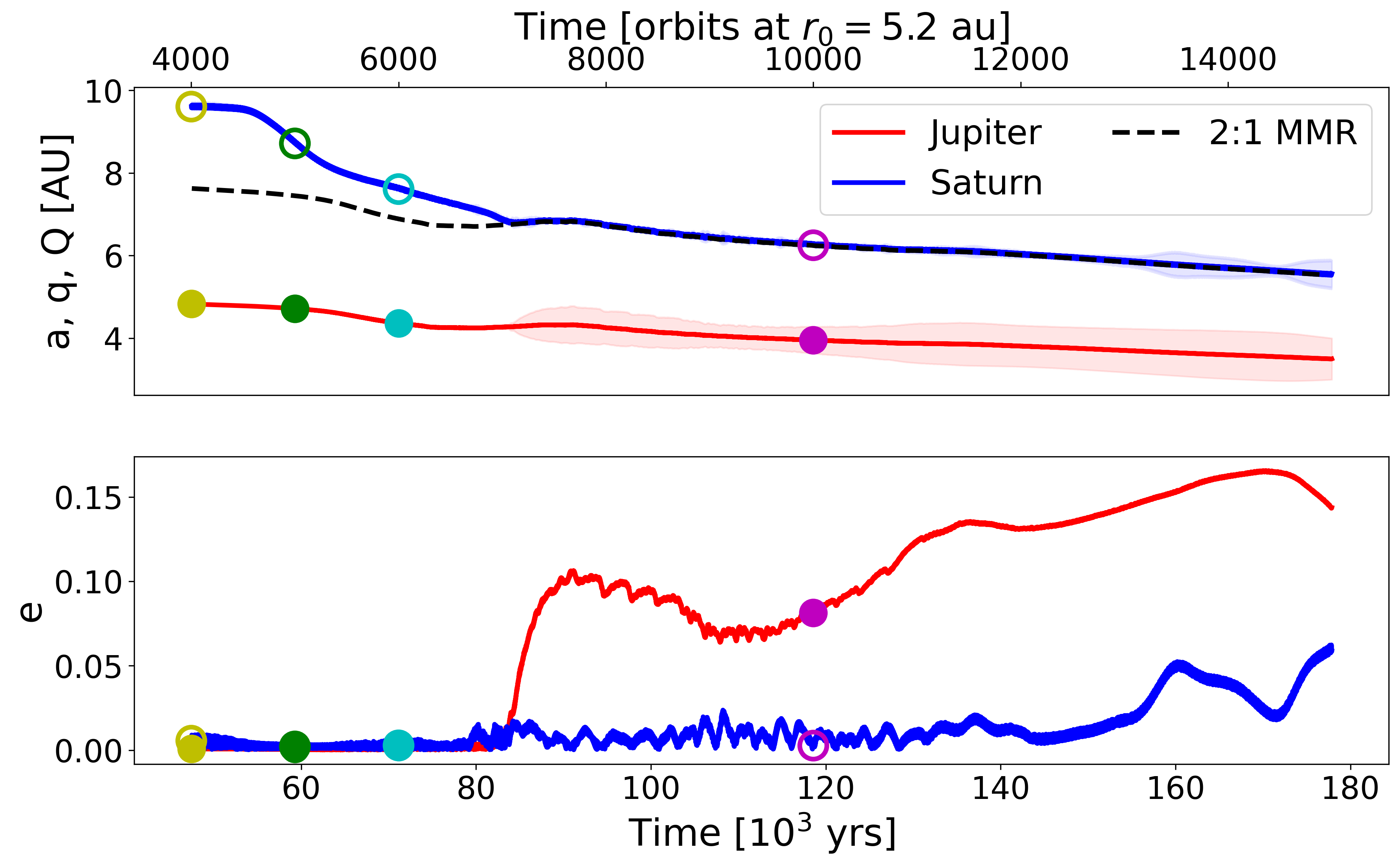}
    \caption{{Orbital parameters' evolution of Jupiter and Saturn in the nominal simulation. The top panel shows the semi-major axes of Jupiter in red and Saturn in blue. {The shaded areas mark the positions of the peri- and apo-centre, $q=a(1-e)$ and $Q=a(1+e)$, of the planets, respectively.} The  black dashed line mark{s} the position of the 2:1 resonance with Jupiter. The bottom panel shows the eccentricities of the planets. At the start of these curves, Jupiter is fully grown, while Saturn reaches its final mass around $72\,700$ years. The coloured markers indicate the times at which the surface density is represented in the four panels of \cref{fig:nom_SigMap}.}}
    \label{fig:nom_OrbPar}
\end{figure}

\begin{figure}
    \centering
    \includegraphics[width=\columnwidth]{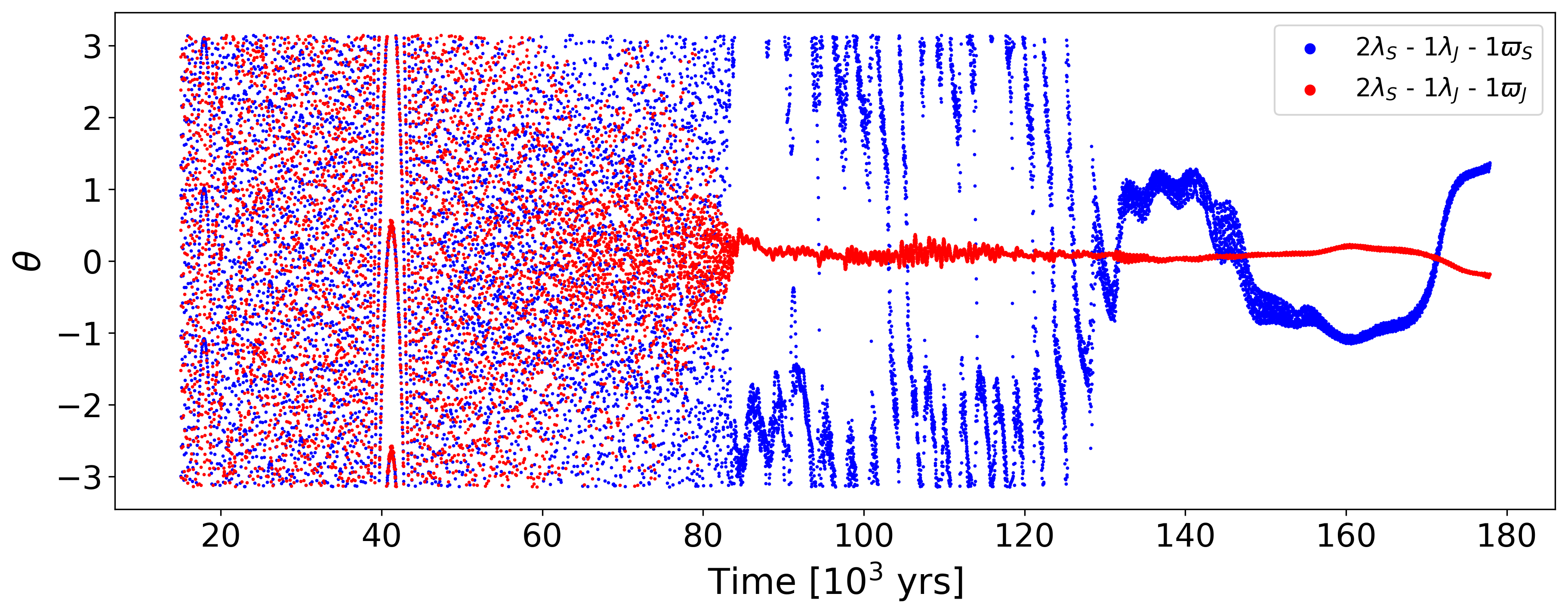}
    \caption{{{Evolution of the resonant angles {$\theta_{J,S}$} of both planets in the nominal simulation. The angle $\lambda_p$ is the mean longitude of planet $p$ and $\varpi_p$ is its longitude for the perihelion.}}}
    \label{fig:nom_resang}
\end{figure}

In our nominal simulation, {Saturn was positioned outside of Jupiter's gap} at a distance of $a_S/a_J = 2$. Panel a of \cref{fig:nom_SigMap} shows the surface density of the disc at the moment of introduction of Saturn. The other panels of \cref{fig:nom_SigMap} show the positions of Jupiter and Saturn in the disc at different moments of the evolution of the system. Panel b shows Saturn, at $45\%$ of its final mass, transitioning from a wake-driven migration to carving its own gap. The gap was fully opened in panel c once the planet was close to its final mass. At last, panel d of \cref{fig:nom_SigMap} shows the disc surface density after $118\,000$ years, when Jupiter and Saturn were migrating together in a common gap. \\
We show the evolution of the semi-major axes and eccentricities of the planets in \cref{fig:nom_OrbPar}. Saturn reached its final mass around $72\,600$ years. The four coloured circles in the figure represent the moments at which the surface density of the disc is shown in the four panels of \cref{fig:nom_SigMap}. \Cref{fig:nom_OrbPar} shows that while growing, Saturn migrated slowly inwards until the pair of planets were locked in the {2:1} MMR. {\Cref{fig:nom_resang} confirms that the two resonant angles of the system are librating and therefore the planets are stably locked into the resonance.} These angles are defined as 
{$$\theta_{J,S} = 2\lambda_S-\lambda_J - \varpi_{J,S}$$}
with $\lambda_p$ and $\varpi_p$ being the mean longitude {and longitude of perihelion} of planet $p$, respectively. \\
When the planets entered into resonance, the eccentricity of Jupiter was excited up to {the order of $\sim 0.1$} while that of Saturn remained {very low.} This is {consistent with $\theta_S$ having a large libration amplitude, while $\theta_J$ remains very close to $0$; however, this is} 
an uncommon result as shown by \cite{michtchenko_dynamic_2008} {regarding the stability of such a configuration}. {Furthermore, until now in the literature, Jupiter-Saturn migration studies had found that in most cases Saturn's eccentricity is higher than that of Jupiter, and this has been seen for the planets locked in the 3:2 MMR \citep[see e.g.][]{masset_reversing_2001,dangelo_outward_2012} as well as in 2:1 MMR \citep{pierens_outward_2014}.} Readers can refer to \cref{sec:discMass} for further information about the eccentricities of the planets.

        \subsection{Disc thickness}

\begin{figure}
    \centering
    \includegraphics[width=\columnwidth]{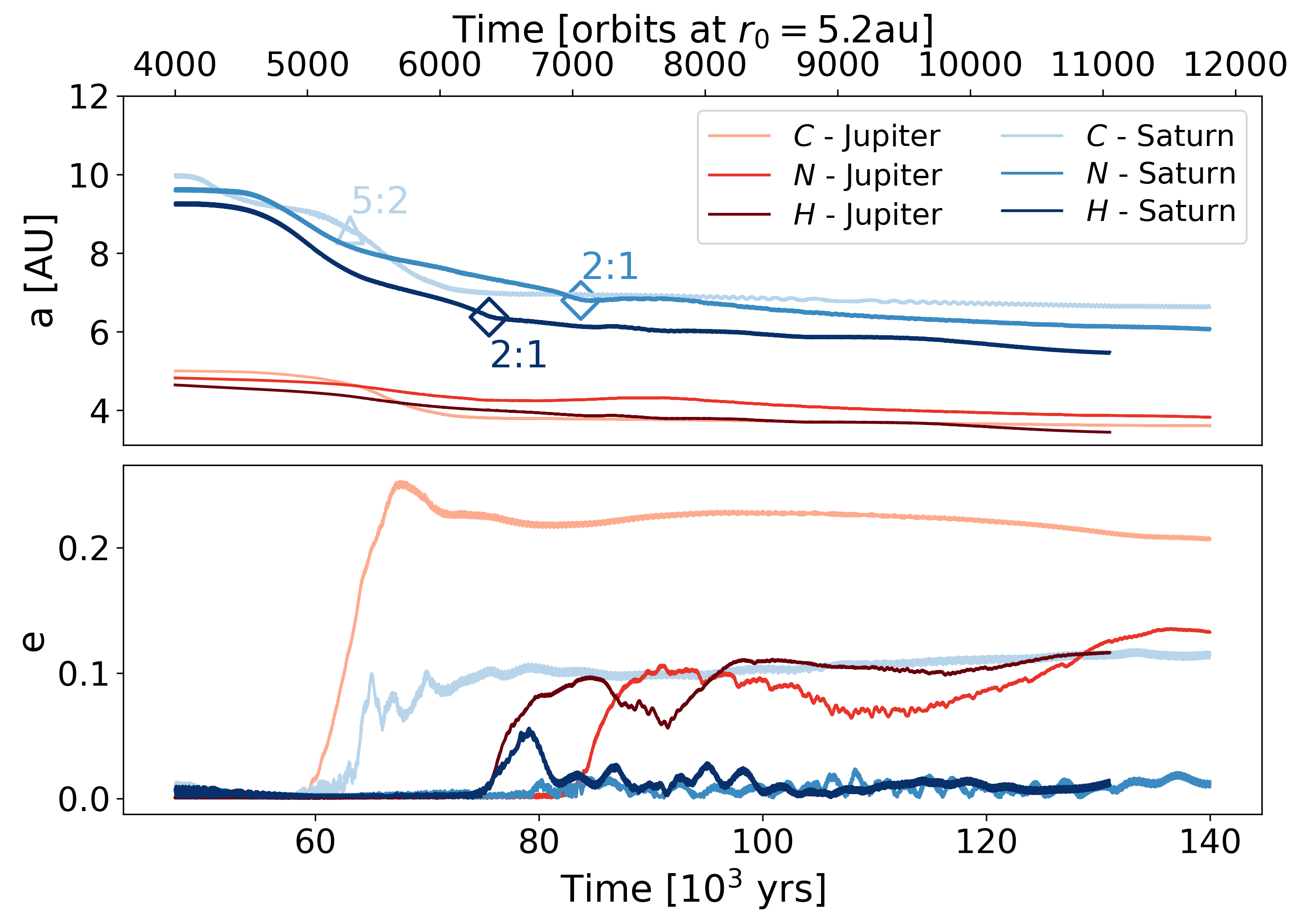}
    \caption{Orbital evolution of Jupiter and Saturn for the {simulations $C$, $N$, and $H$} (see {Table} \ref{tab:simulations}). The top panel shows the semi-major axis of both planets (Jupiter in red, Saturn in blue), and the bottom panel shows their eccentricities. The markers indicate the time at which the planets get locked into a resonance: the triangle marks the 5:2 MMR, and the diamond marks the 2:1 MMR.}
    \label{fig:h_OrbPar}
\end{figure}

\begin{figure}
    \centering
    \includegraphics[width=\columnwidth]{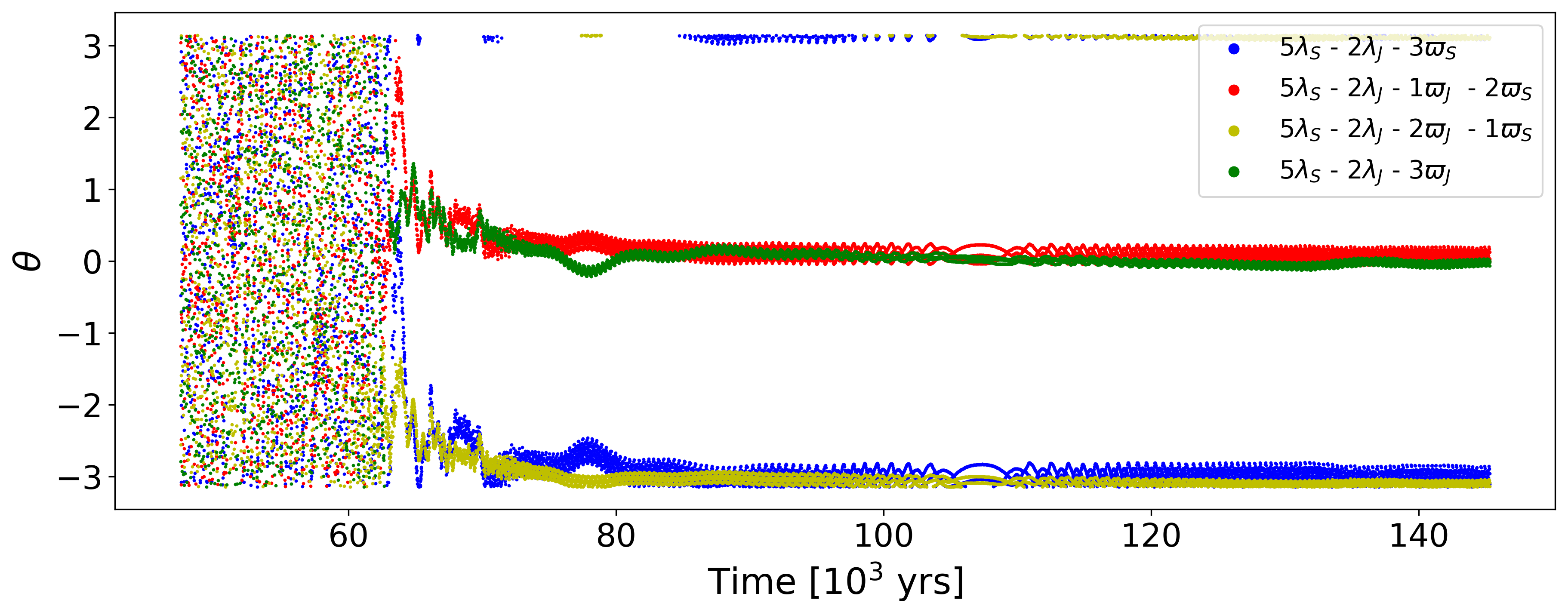}
    \caption{{Evolution of the four resonant angles in {simulation $C$}.}}
    \label{fig:h-_resAng}
\end{figure}

\begin{figure}
    \centering
    \includegraphics[width=\columnwidth]{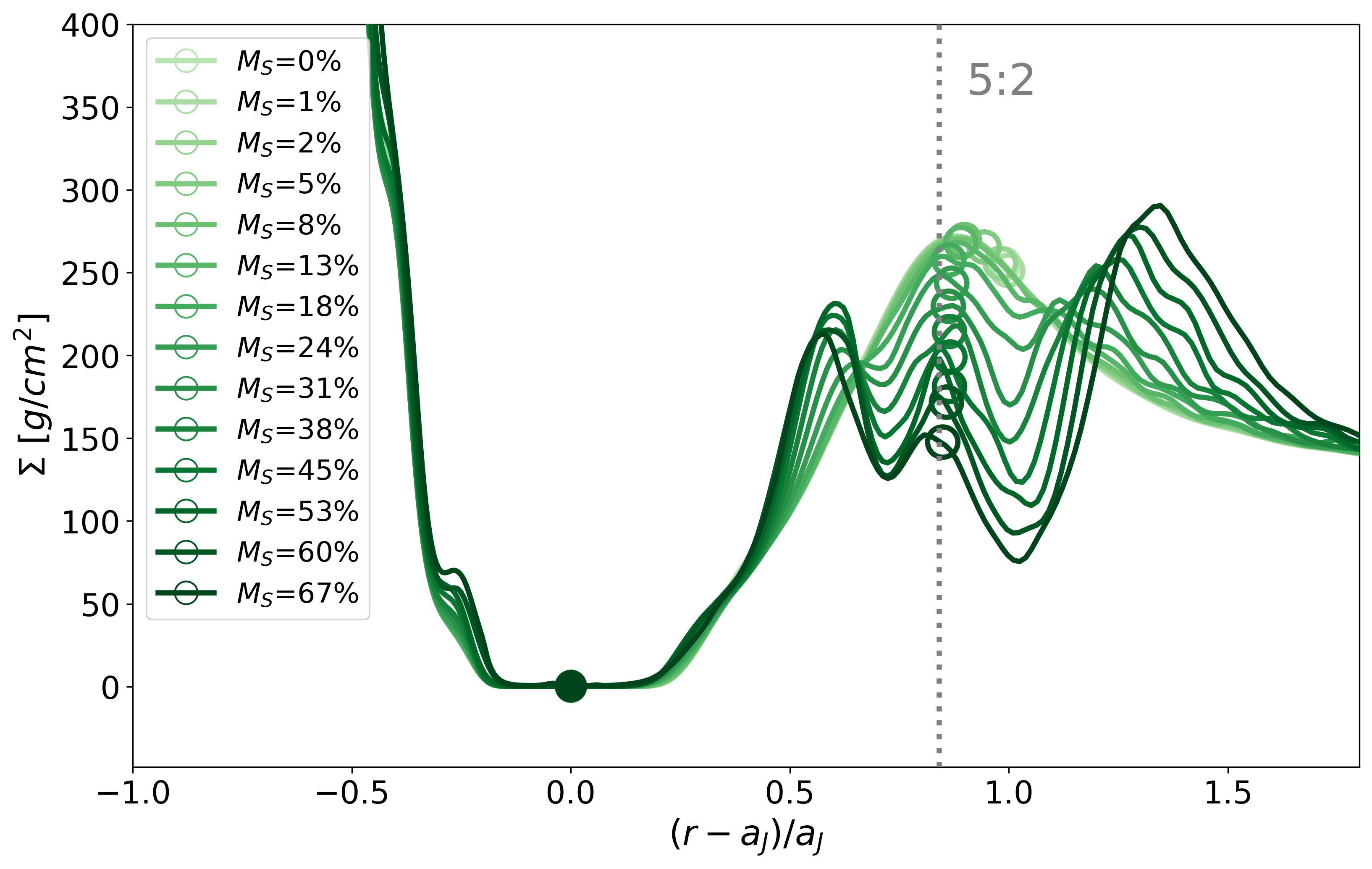}
    \caption{{Radial density profiles of} { simulation $C$} {in the time interval $[45\,500,60\,000]$ years. The green colour gradient corresponds to a time evolution (and therefore also a mass evolution for Saturn) from light to dark. The filled and empty circles indicate the positions of Jupiter and Saturn.}}
    \label{fig:h-_densProf}
\end{figure}

In classical discs, for $\alpha \geq 10^{-3}$, it has been shown that the disc thickness can influence the direction of migration of a pair of Jupiter-Saturn-like planets when they are locked in a 3:2 MMR, \citep{morbidelli_dynamics_2007-1}. {Therefore,} in this section, we show how the result obtained for the nominal simulation varies with the disc thickness.\\
In  \cref{fig:h_OrbPar}, we show the evolution of the orbital parameters of both planets for simulations with a {colder, }thinner disc, {$C$ with $h_0 = 0.035$} and with a {hotter, }thicker disc, {$H$ }with $h_0=0.06$. In the thicker disc, as in the nominal case, the planets are locked into a {2:1} MMR. While in the thinner disc, the planets end up in a {5:2} MMR configuration. \\
{We shall focus on case $C$ and its unusual outcome. We see in \cref{fig:h_OrbPar}}
{that while growing, Saturn migrates slowly inwards until the pair of planets gets locked in a 5:2 MMR. \Cref{fig:h-_resAng} shows the libration of the resonant angles of the system and confirms that the planets are stably locked into a 5:2 resonance.} \\
{To understand further why Saturn's migration slows down before the 5:2 MMR and is captured into this resonance, we show in \cref{fig:h-_densProf} the evolution of the azimuthally averaged density profiles of the disc during the phase of inward migration and growth of Saturn. The profiles are centred on Jupiter's position in order to highlight Saturn's motion with respect to the 5:2 MMR. The position of Jupiter and Saturn are marked with filled and empty circles, respectively. The shape of the gap carved by Jupiter depends on the disc's aspect ratio \citep{crida_width_2006}. Indeed, the thinner the disc, the steeper the density slope is at the edge of the gap {and the wider the gap is. This can result in the creation of a density bump at the outer edge of Jupiter's gap, beyond the 2:1 MMR. Besides, for a fixed $\alpha$, the viscosity is proportional to $h^2$ so that it is {twice as small} in the $C$ case than in the nominal $N$ case. As a consequence, the bump is less quickly and less easily smoothed and eroded. While in case $N$, we did not observe any bump, in the thin disc case $C$ there is a clear bump that can be seen from the lightest green curve in \cref{fig:h-_densProf} corresponding to the disc profile at time $T_{0,S}$. \Cref{fig:h-_densProf} shows that} while migrating in the type I regime, Saturn gets trapped at the density maximum located at the edge of Jupiter's gap. This density bump acts as a 'planet trap' where the steep positive gradient in density leads to a positive co-rotation torque that cancels out the negative total Lindblad torque \citep{masset_surface_2006}. Once trapped, the planet continues to grow until it starts forming its own gap. The trapping of Saturn at the density peak slows down its migration considerably, allowing it to get locked in the 5:2 MMR.} {Finally, we note that the eccentricities of both planets are much higher in the thin disc case.} \\
We find that the disc's aspect ratio has an influence on the resonant configuration of the pair of planets, through the initial gap profile of Jupiter. The planet trap at the edge of Jupiter's gap will determine whether Saturn will be locked in {the 5:2} resonance or migrate further inwards towards the 2:1 MMR. We therefore expect that there is a threshold for the disc height {below} which the planets will end up being in the 5:2 MMR, or in the 2:1 MMR otherwise. \\
The {overall behaviour of} the pair of planets, however, does not differ significantly whether they are in a 5:2 or 2:1 resonance. {Precisely}, we notice the {direction of} the migration of the pair is inwards in both resonant configurations. {In} both cases the orbital configuration is not compact enough for the reversal of the migration direction to {occur} as is typically observed in a 3:2 MMR. A comment on the inward migration speed is subsequently provided in the discussion (see \cref{sec:discussion}).

        \subsection{Disc mass}\label{sec:discMass}

\begin{figure}
    \centering
    \includegraphics[width=\columnwidth]{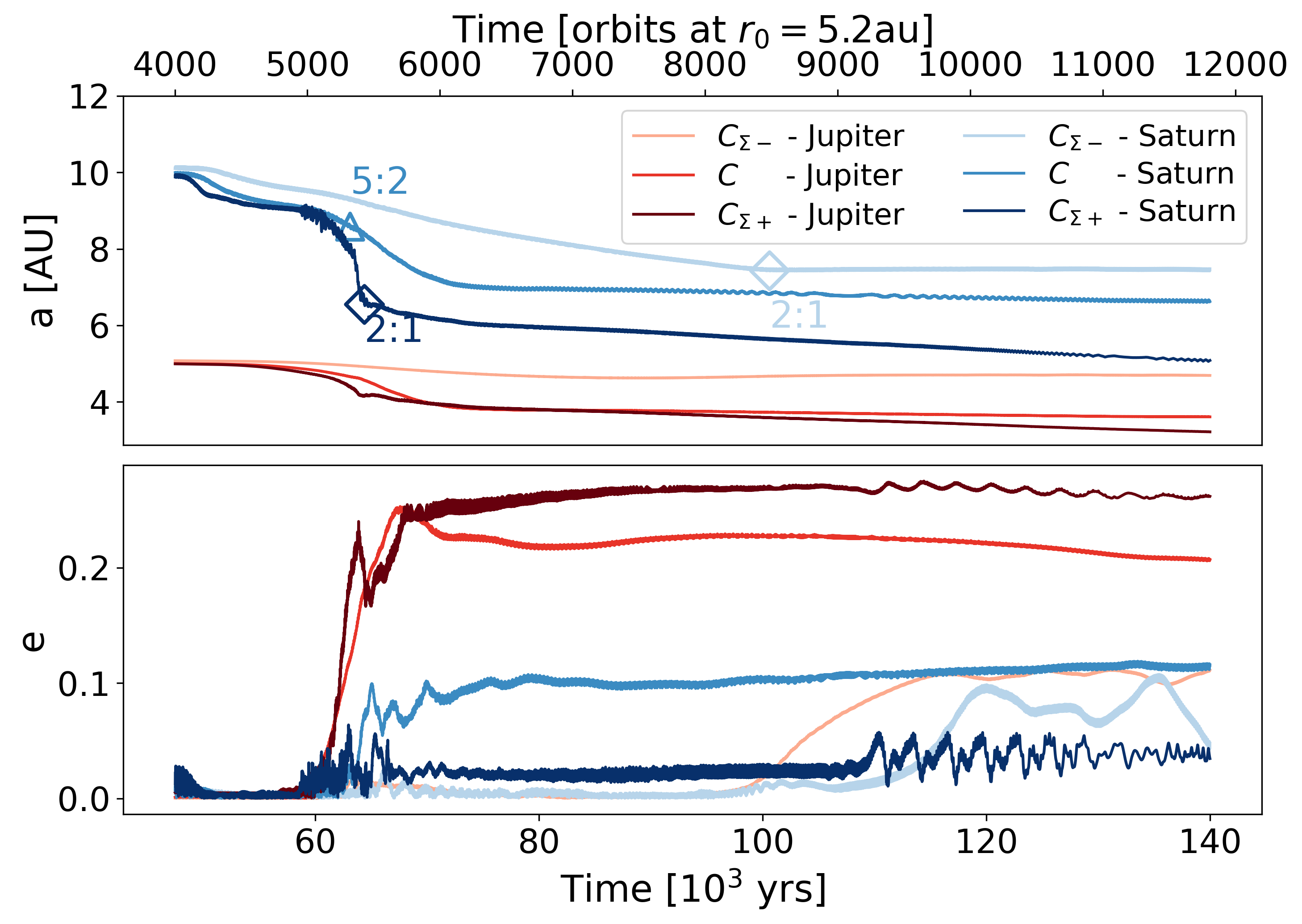}
    \caption{Same as \cref{fig:h_OrbPar}, but for {simulations $C_{\Sigma -}$, $C$, and $C_{\Sigma +}$.}} 
    \label{fig:sig_OrbPar}
\end{figure}

\begin{figure}
    \centering
    \includegraphics[width=\columnwidth]{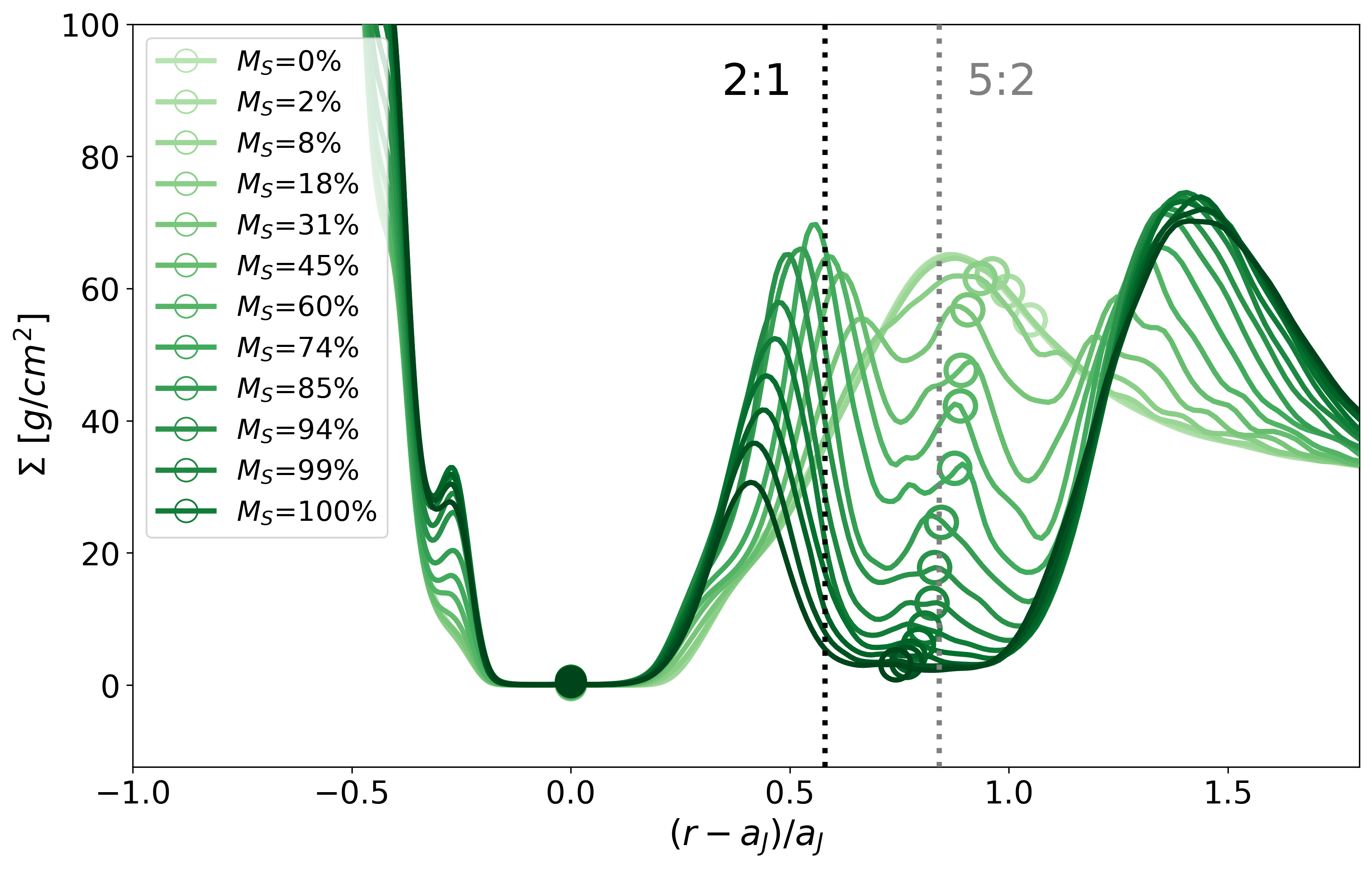}
    \caption{Same as in \cref{fig:h-_densProf}, but for the time interval $[45\,500, 83\,000]$ years, {for {simulation $C_{\Sigma-}$}}.} 
    \label{fig:sig_densProf}
\end{figure}

We are now interested in studying the effects of the disc mass on the migration of the pair of planets. We have run the simulations $N_{\Sigma -}$ and $N_{\Sigma +}$ from Table \ref{tab:simulations} and have found that just as for the nominal simulation, the pair gets locked into a 2:1 MMR. We therefore conclude that the result obtained for the nominal simulation is robust {when the disc mass changes}. We are however interested in exploring the result obtained in the previous section for simulation $C$.\\
Therefore, we explored different disc masses in the case of the thin disc, and studied how robust the 5:2 MMR outcome is. \Cref{fig:sig_OrbPar} shows the evolution of the orbital parameters comparing {the thin disc with nominal mass $C$} with a less massive, {$C_{\Sigma -}$}, and a more massive disc, {$C_{\Sigma +}$}. {Since the shape and depth of {Jupiter's} gap does not depend on the surface density, in all three cases Saturn starts at a similar position with respect to the local density distribution.} In the more massive disc, $C_{\Sigma +}$, Saturn's migration starts similarly {as in $C$} and the planet initially gets captured in the 5:2 resonance before $60\,000$ years. However, {because} the disc {is more massive,} the force it applies to the pair of planets is strong enough to force Saturn out of the 5:2 resonance and make the planet continue its migration inwards until it reaches the 2:1 resonance. Once in this resonance, the pair of planets migrates inwards, significantly faster than in the nominal {disc mass} case.\\
{Since type I migration is proportional to the disc mass {\citep{paardekooper_torque_2011}},} in the less massive disc {$C_{\Sigma -}$}, the migration speed of the Saturn embryo is slower than in case $C$. {As a consequence, Saturn opens its gap before reaching the planet trap.} Instead, the planet enters a regime of type II migration and crosses the 5:2 MMR unperturbed. This is shown in \cref{fig:sig_densProf}.\\
In the {$C_{\Sigma -}$ case}, the eccentricities of Jupiter and Saturn both oscillate around $10\%$. This indicates that the high eccentricity of Jupiter in the other simulations is induced by the gas' potential in the surroundings of the planets. We have tested this hypothesis {in \cref{app:gasRem}}.

        \subsection{Distance between planets}\label{sec:asaj}

\begin{figure}
    \centering
    \includegraphics[width=\columnwidth]{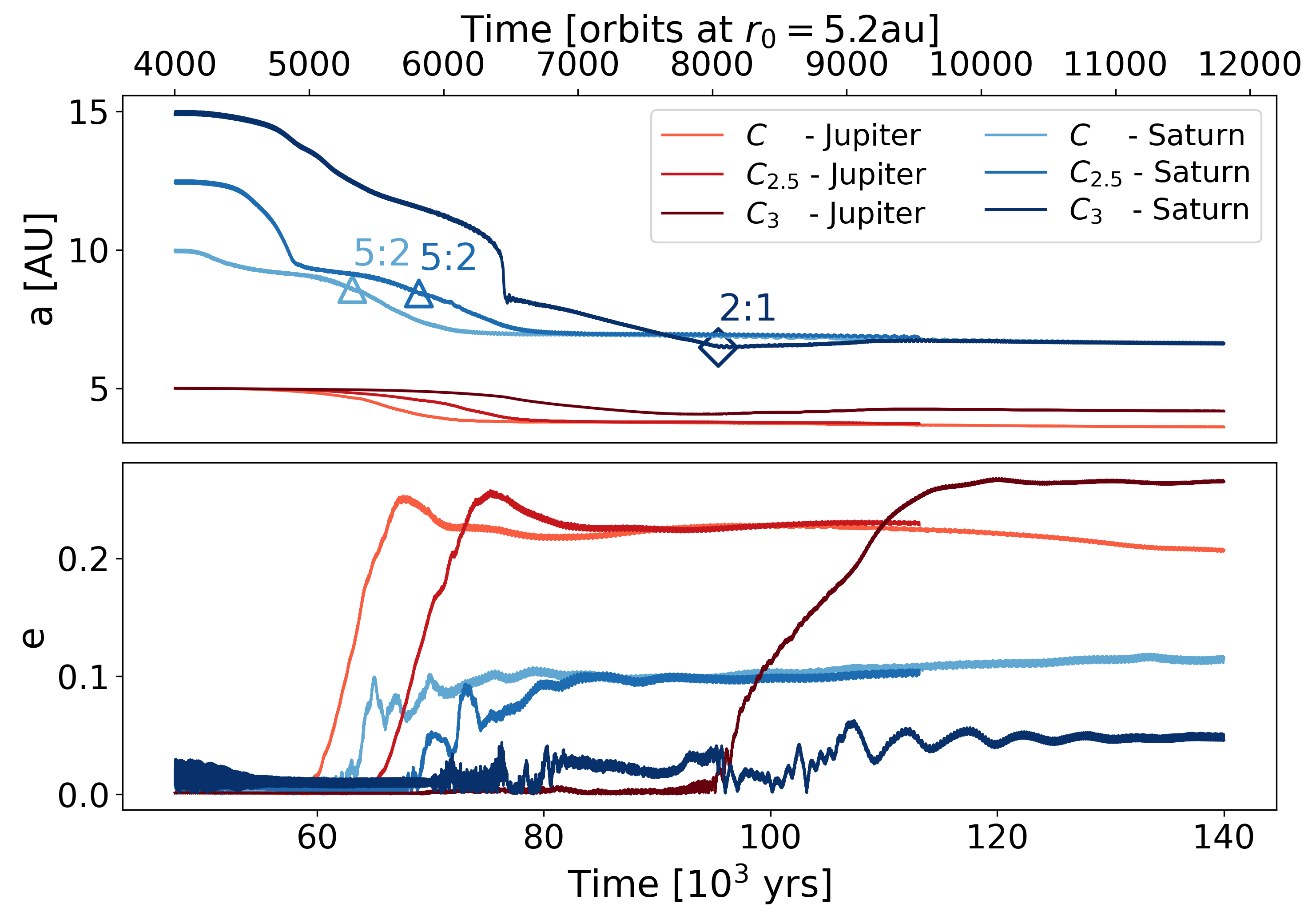}
    \caption{Same as \cref{fig:h_OrbPar}, but for {simulations $C$, $C_{2.5}$, and $C_{3}$.} {Simulation $C_{2.5}$ was stopped at about $110\,000$ years as it overlaps almost perfectly with $C$ in both orbital parameters.}}
    \label{fig:asaj_OrbPar}
\end{figure}

\begin{figure}
    \centering
    \includegraphics[width=\columnwidth]{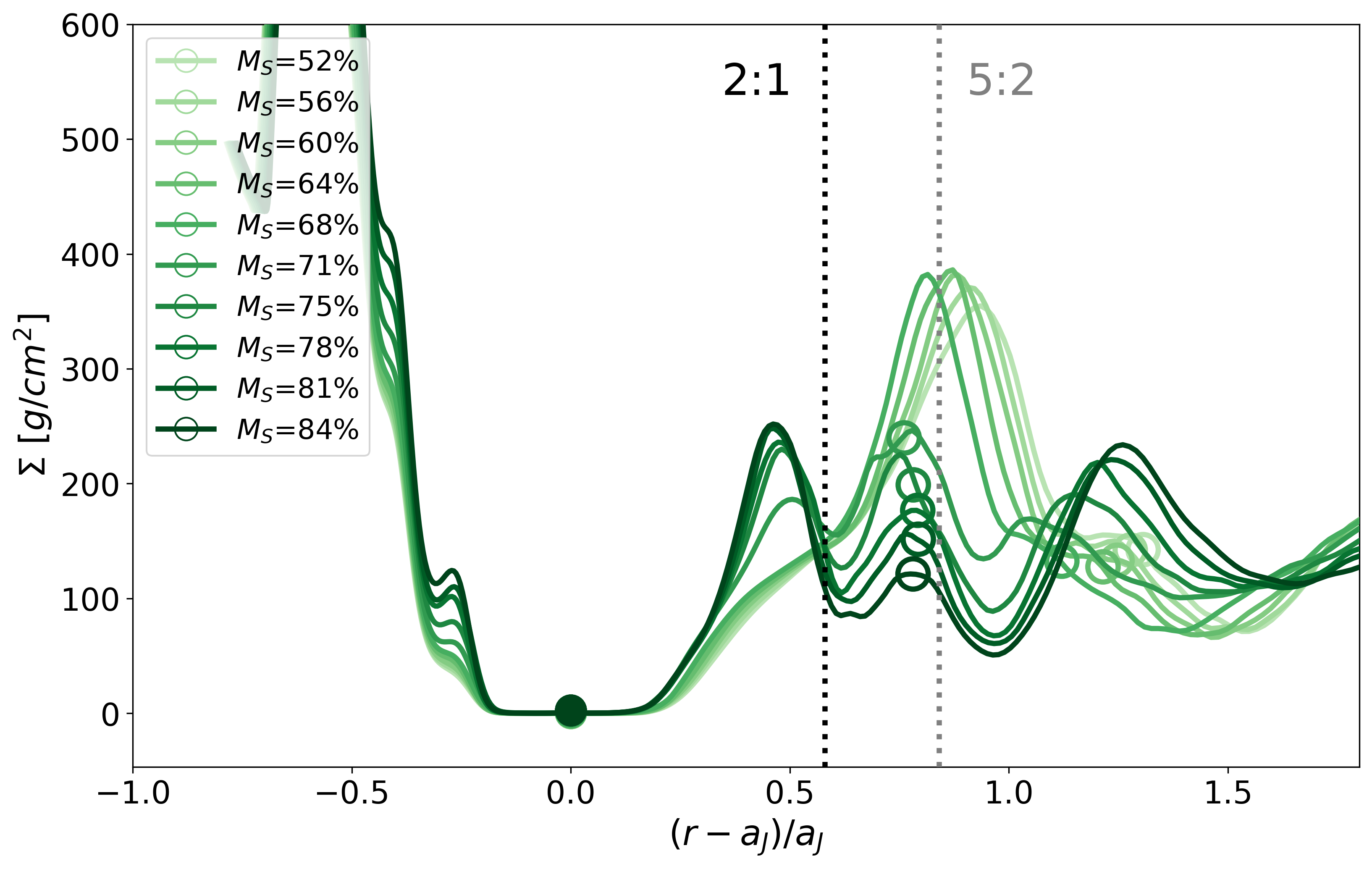}
    \caption{Same as in \cref{fig:h-_densProf}, but for {simulation $C_3$} in the time interval $[75\,000, 80\,000]$ years. It is important to note the jump of Saturn between masses 68\% ($r = 2.12a_J$) and 71\% ($r=1.76a_J$) in 100 orbits at $r_0$.} 
    \label{fig:asaj_SurfDens}
\end{figure}

The dependence of the outcome of the migration of Jupiter and Saturn on their initial distance separation has been shown in \cite{chametla_capture_2020}. In this section, {we investigate how the initial position of Saturn with respect to the pressure bump present in thin disc case $C$ may affect the final MMR configuration of the pair.
As there is no pressure bump in nominal case $N$, we expect the planets to end in the 2:1 MMR in any case (provided it starts outside of the 2:1 and not too far away) and therefore we only explore this parameter in the cold disc case.} \\
We ran simulations in which we placed Saturn at a distance $a_S/a_J = 2.5$ and $a_S/a_J = 3$. As explained in \cref{sec:units}, the important quantity is the ratio of the semi-major axes (or equivalently of the orbital periods), not the difference\footnote{Obviously, a 1 {au} separation does not correspond to the same dynamical situation at 1 {au} and at 10 {au}}. We recall that the growth of Saturn is set to last 800 initial orbital periods, which entails the growth time of the planet being longer the further away it is placed in the disc. The consequence of this is that the planet remains longer (in physical time, not in orbital periods) in a fast migration regime compared to the nominal case. If the growth time is shorter, Saturn carves its gap before reaching Jupiter and both planets migrate independently. \Cref{fig:asaj_OrbPar} shows the results of these simulations. \\
 {Case $C_{2.5}$} is very similar to the {$C$ simulation}; both planets are locked in the 5:2 resonance and their final eccentricities reach the same values. In the {$C_3$ case}, the pair of planets end in the 2:1 configuration; this is due to Saturn experiencing a fast inward migration episode between $75\,000$ and $80\,000$ years. \Cref{fig:asaj_SurfDens} shows the density profiles of this simulation at the time of this episode. The planet slowly migrates inwards within the low mass migration regime until {it starts carving its gap near} the edge of Jupiter's gap. {This pushes gas between the two planets creating an even stronger density peak than in the case of simulation $C$.} That steep density gradient combined with a co-orbital mass deficit has positive feedback on the inward migration of Saturn, triggering a short episode of runaway migration. As a consequence, Saturn crosses the 5:2 MMR and continues its migration path until it encounters the 2:1 MMR. We {recall} that Saturn does not accrete gas from the disc while it grows. {If this had been the case, this} could have a consequence on the size of the density peak located between the two planets and therefore on the episode of the fast migration of Saturn. \\
Finally, if Saturn is formed further away in the disc ($a_S \gtrsim 5 a_J$), it would not migrate fast enough to catch up to Jupiter and both planets would have independent migrations. {We also assess in \cref{app:outward} the case in which Saturn would form {on the inside of} the 2:1 MMR, at an initial distance of $a_S/a_J=1.4$. In this case only, we find Saturn and Jupiter locked in a 3:2 MMR and the pair migrates outwards. However this scenario requires that Saturn form inside Jupiter's gap and this is therefore unlikely.}

\section{Why the 2:1 MMR is never crossed} \label{sec:criterion}

We have seen so far that in low-viscosity discs, it seems that Saturn never crosses the 2:1 resonance. Indeed the planets are never in the 3:2 MMR and as a consequence the pair never migrates outwards. This is in contrast with results found in the literature that consider an $\alpha$ viscosity parameter no less than $10^{-3}$ {\citep[e.g. ][]{masset_reversing_2001,morbidelli_dynamics_2007-1,dangelo_outward_2012}}. Therefore, a relevant question to {assess} in this work is why the 2:1 resonance is never crossed at low viscosity{.} In this section, we answer this question using a semi-analytical method {and generalise this result}. \\

\subsection{Resonance crossing criterion}\label{subsec:criterion}

\begin{figure}
    \centering
    \includegraphics[width=\columnwidth]{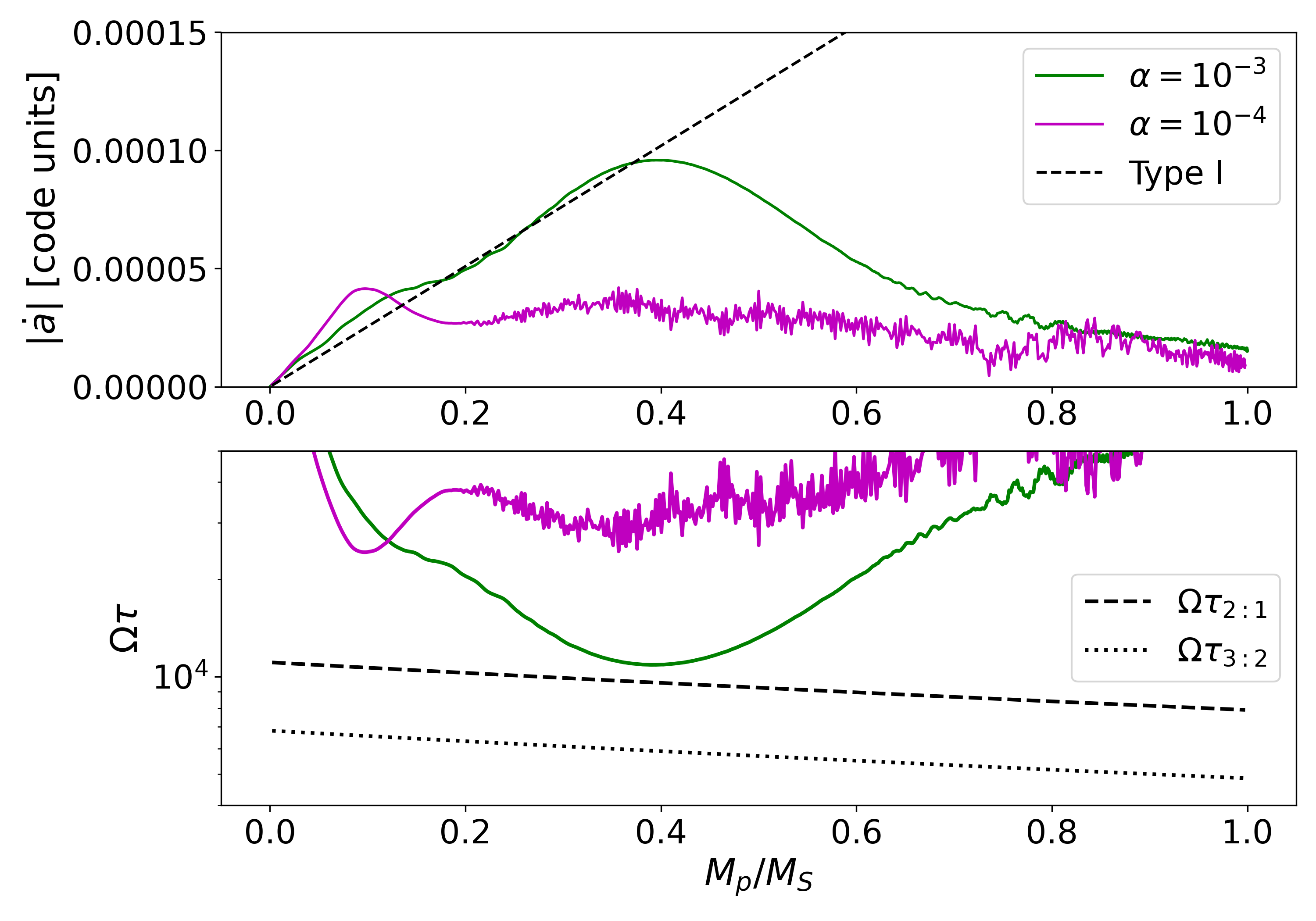}
    \caption{{Migration speed and timescale of a growing planet in different viscous discs.}
    Top panel: Migration speed in absolute value of a planet growing from zero to Saturn's mass, $M_S$, in $800$ initial orbital periods. The dashed black line shows the type I migration speed from \cite{tanaka_threedimensional_2002} (at $r=r_0$ constant). Bottom panel: Migration timescale $\tau = {a}/{|\dot a|}$ normalised by the angular velocity of the planet. The black dashed and dotted lines represent the critical migration timescale, calculated with \cref{eq:tauCrit}, beyond which the planet would cross the 2:1 or 3:2 resonance, respectively. These simulations were run for {viscosities corresponding to $\alpha=10^{-4}$ and $10^{-3}$}. The migration timescales plotted in this figure were smoothed by means of a sliding window average with a size of then and $50$ points (which represents a time interval of about $100$ and $600$ years) for the case $\alpha=10^{-3}$ and $\alpha=10^{-4}$, respectively.}
    \label{fig:migRate}
\end{figure}

\begin{figure}
    \centering
    \includegraphics[width=\columnwidth]{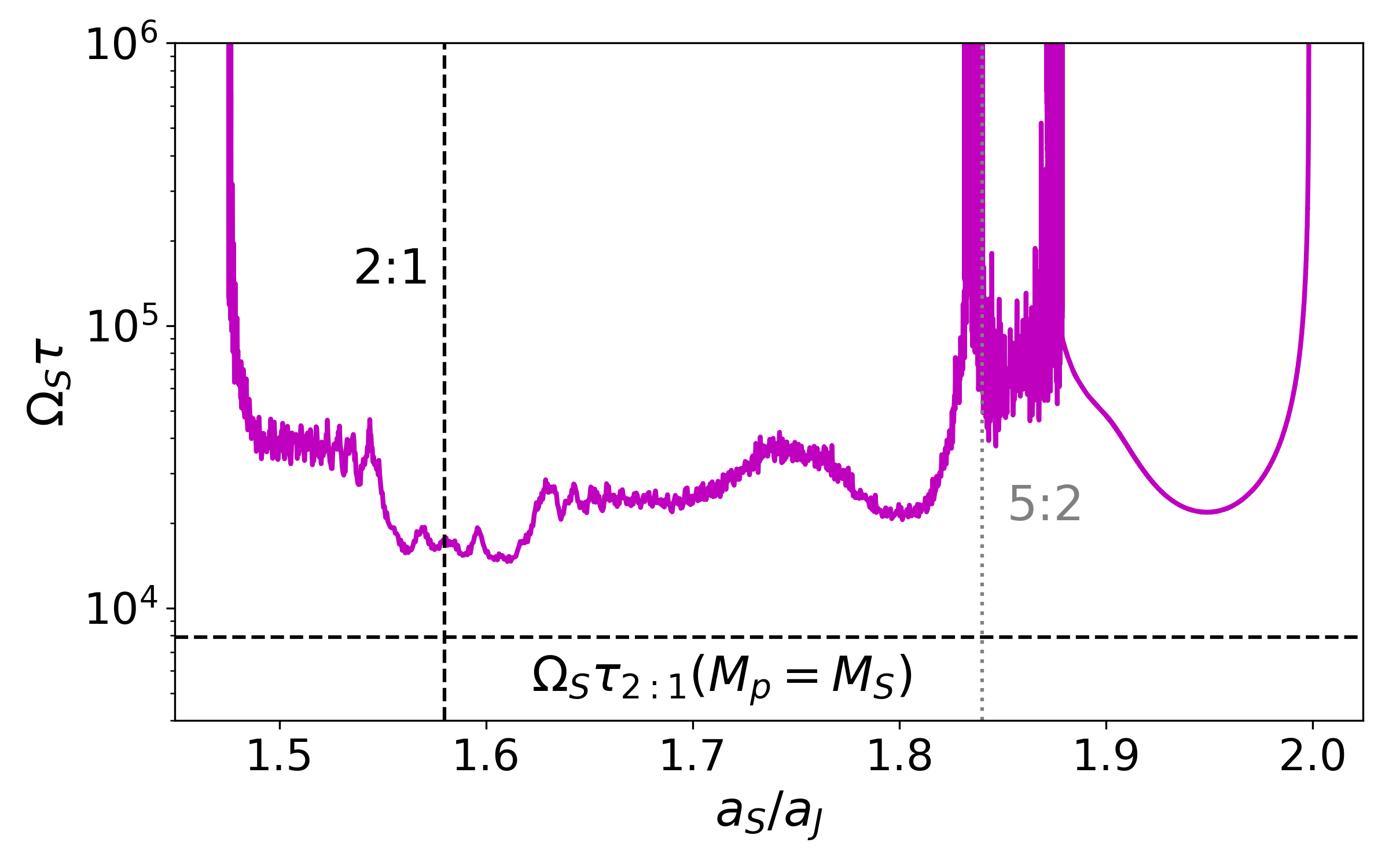}
    \caption{Relative migration timescale $\tau = \frac{a_S/a_J}{d(a_S/a_J)/dt}$ in a simulation where Saturn migrates in the {cold disc} perturbed by Jupiter, without feeling Jupiter's gravity. The normalised timescale is plotted as a function of $a_S/a_J$, and the vertical black dashed line marks the location of the 2:1 MMR, while the grey dotted vertical line marks the position of the 5:2 MMR. The horizontal dashed line marks the critical migration timescale, calculated from \cref{eq:tauCrit} (for $M_p=M_S$), above which the planets should not cross the 2:1 resonance. The migration timescale was smoothed by means of a sliding window average with a size of $100$ points (which represents a time interval of about $12\,000$ years), in order to suitably reduce the noise.}
    \label{fig:ResCross}
\end{figure}

The crossing of a resonance is dictated by the relative migration speed between the two bodies and the libration period of their resonant angle. This has been studied by \cite{dangelo_outward_2012} and further developed in \cite{batygin_capture_2015}. The later paper has shown analytically that the outer body gets locked into a resonance if the libration period of the resonant angle is shorter than the characteristic {timescale} of convergence. Evidently this criterion for resonance locking depends on the migration rate of both planets. \\
If we know the analytical migration rate of the planets, we can derive a fully analytical method to understand the resonance crossing. In \cite{batygin_capture_2015}, Section 3.1.2 (i), the author derives the criterion specifically applied to an inner planet migrating in a type II regime and an outer, less massive planet following a type I migration (cf. {their} equation 50). However, we find that this derivation is based on assumptions that are too restrictive to be applied to our case. Firstly, the author assumes that most of the planetary mass is contained within the inner body and therefore neglects the mass of the outer planet in his derivation. In the case of Jupiter and Saturn with a mass ratio of $M_J/M_S \simeq 3$, it can be argued that this assumption is no longer valid. But most importantly, this criterion was derived using the analytical expression of type I migration from \cite{tanaka_threedimensional_2002} for the outer planet. 
However, type I migration is valid only up to a certain planetary mass. This transition mass depends on the viscosity of the disc. \\
To illustrate this point, we ran simulations of a planet growing from zero mass to $M_p = M_S$ in $800$ initial orbital periods for two viscosity parameters $\alpha= 10^{-4}$ and $10^{-3}${ as well as the aspect ratio of $h_0=0.035$} and we studied its migration as a function of its mass. The top panel of \cref{fig:migRate} shows the migration speed of the planet in absolute value, in addition to which we have superposed the analytical form of type I migration from \cite{tanaka_threedimensional_2002} (which we recall is independent of the viscosity) given for a fixed $r=r_0$. With this figure, we show that the planet in a low-viscosity disc departs from the type I migration regime before reaching $10M_\oplus$ ({while not before} $40M_\oplus$ for $\alpha=10^{-3}$). This is in agreement with the literature \citep{mcnally_migrating_2019, fung_save_2017}. As a consequence, using this analytical expression as the migration rate of the outer planet is not applicable in our configuration and we are therefore unable to use equation 50 of \cite{batygin_capture_2015}. \\
Alternatively, in order to understand the locking of Saturn and Jupiter in the 2:1 resonance in our simulations, we used the general form of the criterion derived in \cite{batygin_capture_2015} (cf. their equation 44). We now assume that the relative migration rate between both planets is dominated by Saturn's migration, thus neglecting Jupiter's migration rate. We rewrite the criterion for resonance crossing as a critical migration timescale below which the outer planet would cross the resonance $k:k-1$, expressed as
\begin{align}\label{eq:tauCrit}
\Omega_S\,\tau_{crit} \simeq \frac{2\pi\, k/(k-1)}{4(3)^{2/3}[k(k-1)^2]^{1/9}}\left(\frac{f_{res}^{(1)}(M_J+M_S)}{M_\star}\right)^{-4/3}
\end{align}
where $\tau = \frac{a}{|\dot a|}$ is normalised by the angular velocity of the outer planet, and $f_{res}^{(1)}$ is a dimensionless constant (that can be found in \cite{deck_first-order_2013}). We note that this criterion applies only to first order resonances. \\
Since we do not have an analytical expression for the migration of a planet in a low-viscosity disc, we used the migration timescales obtained by {the simulations presented in \cref{fig:migRate}} and compared it with the critical migration timescale for resonances 2:1 and 3:2 calculated from \cref{eq:tauCrit}. We show in the bottom panel of \cref{fig:migRate} that, in the case of $\alpha = 10^{-3}$, the minimal migration timescale reached by the planet is close enough to the critical value to cross the 2:1 resonance, while it remains far from that of the 3:2 resonance. This illustrates the well-known result obtained for Jupiter and Saturn migration in 'classical' viscous discs, that is the pair locked in the 3:2 MMR. In the low-viscosity case, however, it is clear that the migration timescale remains much higher than the timescale required to cross the 2:1 MMR. This seems to indicate that a planet growing and migrating in such a disc never migrates fast enough to cross the 2:1 MMR. \\
In the simulation presented in \cref{fig:migRate}, the planet grows in an unperturbed disc. In fact, our study concerns the case of Saturn migrating in a disc where Jupiter has already formed a gap. Therefore, we ran a simulation in which we let Saturn migrate in the disc perturbed by the presence of Jupiter, but without it feeling the gravitational potential of Jupiter\footnote{Nor the indirect term associated with it, see \cite[and in prep.]{Crida+2022_SF2A}}. This allowed us to take into account the presence of Jupiter's gap in the migration of Saturn as this one grew. \Cref{fig:ResCross} shows the relative migration timescale (that is the variation timescale of the ratio of the semi-major axes $a_S/a_J$) in the perturbed disc as a function of the ratio of semi-major axes between the two planets. This figure confirms that the relative migration timescale at the time when the planets approach the 2:1 MMR is much higher than the critical value to cross that resonance. We recall that, in this simulation, Saturn does not feel the presence of Jupiter and therefore does not remain locked in the said resonance. {We also checked that the same is true for our nominal aspect ratio of $h_0=0.05$.}\\
Incidentally, we also note in \cref{fig:ResCross} that Saturn stops at the planet trap just outside the 5:2 resonance, as has already been discussed in \cref{fig:h-_densProf}. When it leaves the planet trap because it starts opening a gap, it approaches the 5:2 resonance very slowly. In this simulation, Jupiter is temporarily caught in this resonance with Saturn (Saturn does not feel Jupiter, but Jupiter feels Saturn such that we recovered the same response in Jupiter's eccentricity as in {simulation $C$}). However, the disc eventually pushes Saturn across the 5:2 resonance and convergent migration resumes.

        \subsection{{Changing the outer planet's mass}}\label{sec:Mp}

\begin{figure}
    \centering
    \includegraphics[width=\columnwidth]{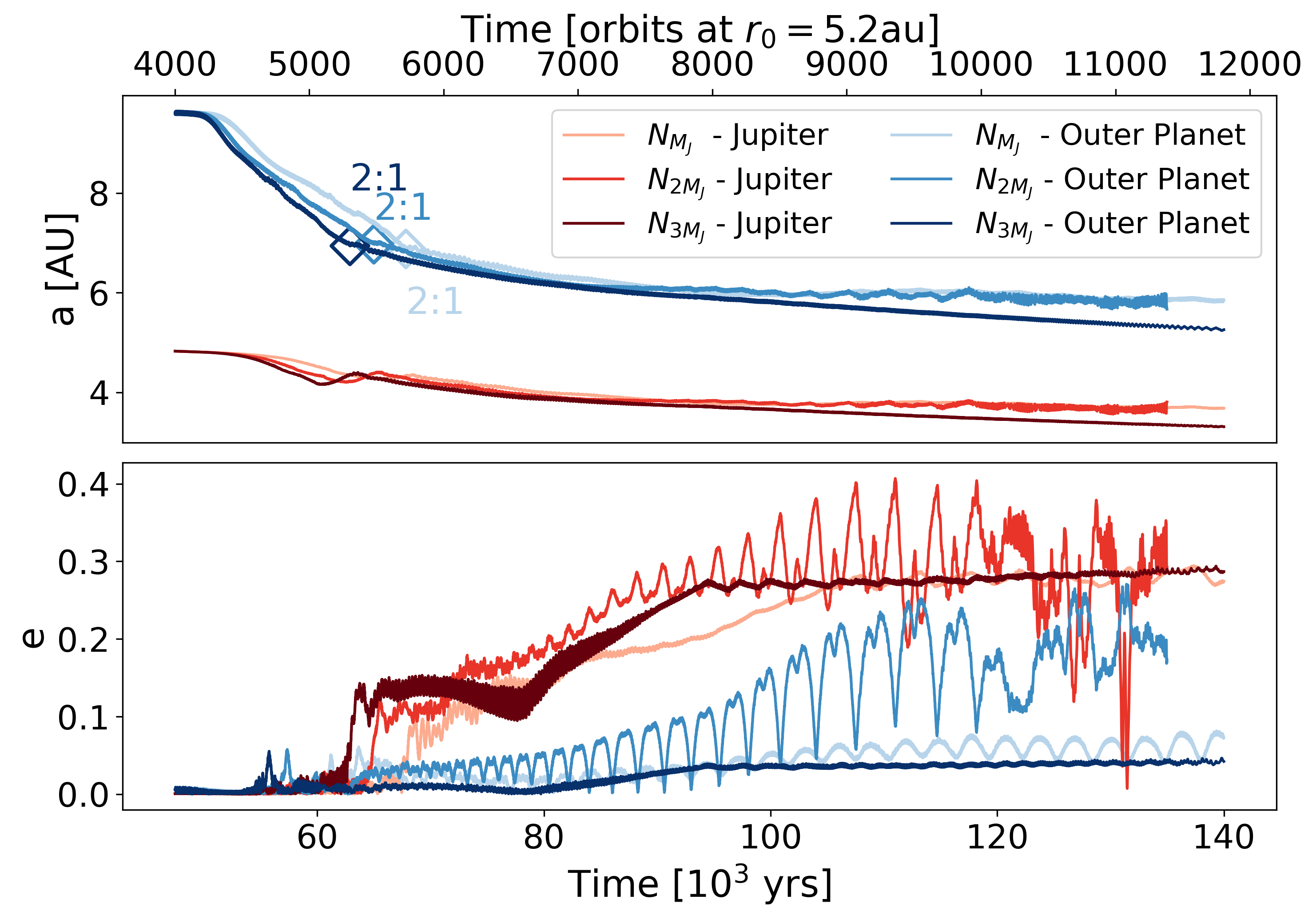}
    \caption{Same as \cref{fig:h_OrbPar}, but for the simulations $N_{M_J}$, $N_{2M_J}$, and $N_{3M_J}$. We note that {$N_{2M_J}$ becomes unstable shortly after $130\,000$ years.}}  
    \label{fig:masses}
\end{figure}

From the analysis performed in \cref{subsec:criterion}, we show that the outer planet does not cross the 2:1 MMR in the case of a low-viscosity disc independently of its mass{, since the peak migration speed is reached way before Saturn's mass}. In this section, we further expand our study by considering a range of masses for the outer planet. The inner planet remains a Jovian mass planet. \\
We show the results of our simulations in \cref{fig:masses} for the following masses of the outer planet $M_p \in [M_J, 2M_J, 3M_J]$ {and $h_0=0.05$}. As expected in all of these cases, the outer planet never crosses the 2:1 MMR with the inner planet. We note, however, that some of these systems{ are not stable} in the long term due to close encounters between the planets. \\ 
In classically viscous discs, the outcome of the migration of pairs of giant planets often required some conditions on the mass ratio of the planets (e.g. Masset and Snellgrove mechanism). This work shows that in a low-viscosity disc, the pair gets {mostly} locked in a 2:1 MMR independently of the mass of the outer planet and this does not require the inner planet to be more massive.

\section{Discussion and conclusions}\label{sec:discussion}

In this paper, we have explored the dynamics of {pairs} of giant planets in low-viscosity discs. We ran two-dimensional hydrodynamical simulations, which allowed us to perform an extensive parameter exploration over long {timescales}. \\
Our results show that the planets {primarily get locked in} a 2:1 MMR. {In some cases, we have found the planets in the higher order resonance 5:2.} In both cases, we do not observe outward migration. In fact, we find that in most cases the planets migrate  inwards, with Jupiter migrating at a speed of the order of {a few} astronomical units per million years. More precisely, {in nominal simulation $N$, we find that the Jupiter-Saturn pair migrates inwards at a similar speed as Jupiter alone, with a speed of the order of $6 \text{{au}}/10^6\text{yr}$. While in the thin disc case, $C$,} once in resonance, Jupiter migrates at a speed of $2\text{{au}}/10^6\text{yr}$; in comparison, the migration speed of Jupiter alone in the {cold} disc is about $10 \text{{au}}/10^6\text{yr}$. Our results therefore show that having a pair of giant planets {does not stop inward migration, but it can in some cases} slow down the migration speed with respect to a single planet. \\
We have shown that the planets never cross the 2:1 MMR, which is at odds with respect to higher viscosity discs. In order to explain our results, we have  used an analytical criterion for resonance crossing based on the work of \cite{batygin_capture_2015}. We show that a planet growing and migrating in a low-viscosity disc never reaches the migration speed required to cross the 2:1 MMR {with Jupiter}.\\
The only case in which outward migration is observed (and therefore is compatible with the Grand Tack model) is a scenario where Saturn would form inside the 2:1 resonance and get locked in the 3:2 {MMR}. However, owing to the large width of Jupiter's gap, this scenario, implying that Saturn forms inside this gap, seems unlikely.
{Thus we conclude the Grand Tack to be almost impossible with $\alpha \lesssim 10^{-4}$. This does not rule out a low-viscosity proto-solar nebula, though, since alternative explanations have been proposed for the small mass of Mars and the depletion of the asteroid belt \citep[e.g.][]{Drazkowska+2016, clement_mars_2018,Clement_terrestrial_2019,nesvorny_role_2021,Izidoro+2022_LPI,Morbidelli+2022}, so the Grand Tack may well not have happened after all.}\\
{As for the Nice model, our results show that if the proto-solar nebula had $\alpha\lesssim 10^{-4}$, the Saturn/Jupiter period ratio would not be less than two after the gas disc phase, in contrast to most of the situations studied in the literature \citep[e.g.][]{tsiganis_origin_2005,morbidelli_dynamics_2007}.
However, the case of Jupiter and Saturn in the 2:1 MMR has been explored by \cite{nesvorny_statistical_2012} and \cite{Deienno+2017}, who found that this configuration is not incompatible with a global instability of the giant planets that would bring them to their current orbits, under some conditions. More recently, \cite{clement_born_2021,clement_born_2021-1} argue that Jupiter and Saturn in the 2:1 MMR with significant eccentricities (more than 0.05 and up to 0.1 and 0.25) are favourable initial conditions for a global 'Nice-model-like' instability. 
Our results support such a setup, although more work is needed to connect our final configuration to their initial conditions (see for instance \cref{app:gasRem}).}\\
{Outside of the Solar System,} we remark that the only system with two giant planets observed in a {protoplanetary disc, to date}, namely PDS 70, occurs to be close to the 2:1 resonance \citep{haffert_two_2019}. \cite{bae_ideal_2019} have shown numerically that a PDS 70-like system initially positioned in the 2:1 MMR, in a classical viscous disc, remains dynamically stable over million-year timescales. However, their study does not address how the two planets have reached such a resonant configuration. Our results suggest that a low viscosity may explain their capture. Simulating the specific case of a PDS 70-like system is beyond the scope of this paper and will be the topic of a future work. \\
We also remark that the two-dimensional low-viscosity discs that we have considered in this paper do not have an accretion rate onto the star compatible with observed values. In a forthcoming study, we will use a new paradigm for the theoretical modelling of discs where the accretion onto the central star occurs in superficial layers while the mid-plane has a close to zero viscosity \citep{lega_migration_2022}. In such discs, the migration of a single giant planet differs from the classical type II migration regime and depends on the thickness of the accretion layer. It is therefore interesting to extend this study to a pair of giant planets.

\section*{Acknowledgements}%
\label{sec:acknowledgements}%

We thank the reviewer for comments that helped in improving the manuscript. We thank Fr\'ed\'eric Masset, Gabriele Pichierri and Michiel Lambrechts for enriching discussions.
We acknowledge support by DFG-ANR supported GEPARD project
(ANR-18-CE92-0044 DFG: KL 650/31-1). We also acknowledge HPC resources from GENCI DARI n.A0120407233 and from "Mesocentre SIGAMM" hosted by Observatoire de la C\^ote d'Azur. We wish to thank Alain Miniussi for maintenance and re-factorisation of the code FARGOCA.

  \bibliographystyle{aux/publisher}

\bibliography{aux/bib}

\clearpage

\appendix

\section{Viscosity and resolution}\label{app:Resolution}

\begin{figure}[h]
    \centering
    \includegraphics[width=\columnwidth]{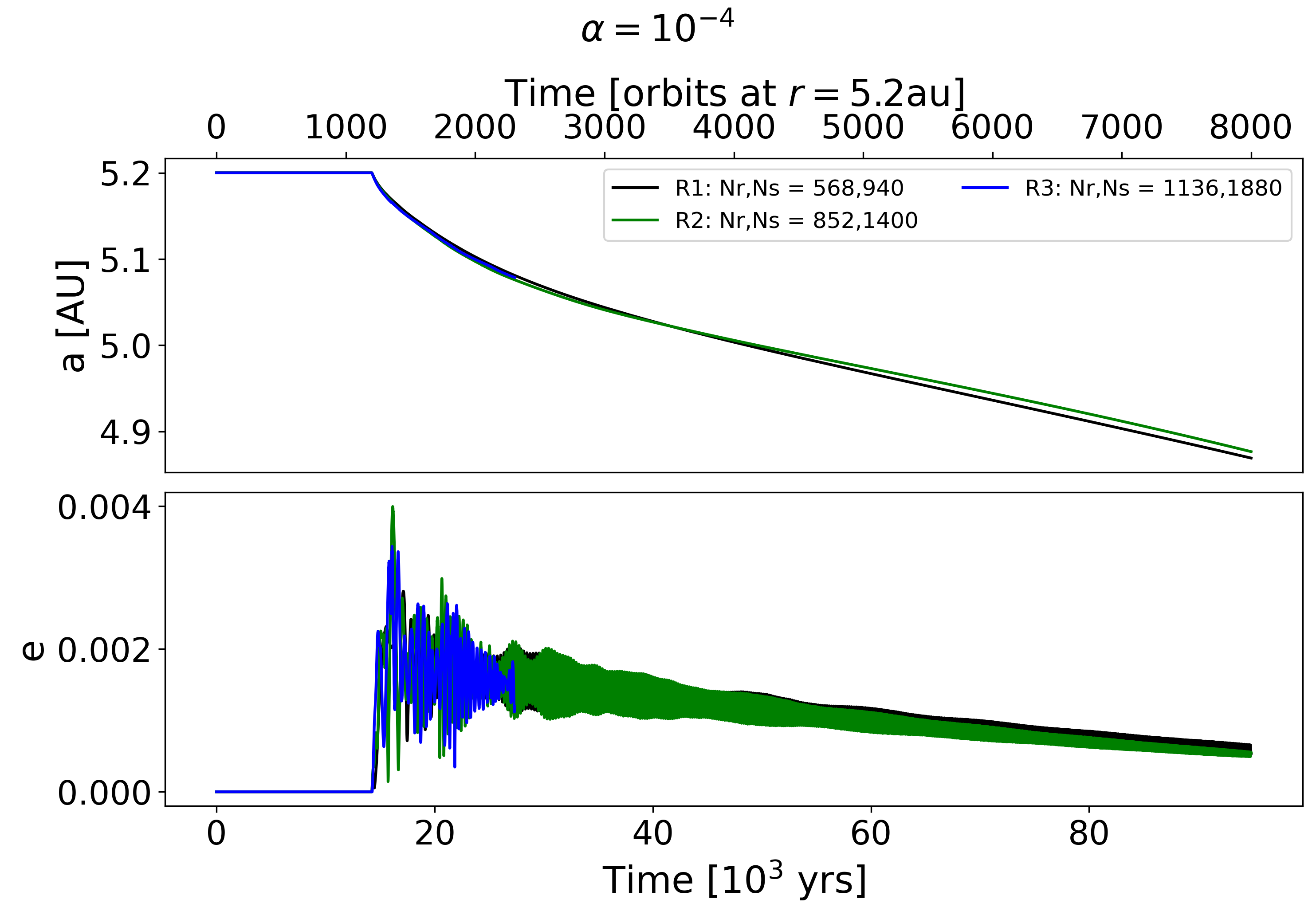}
    \caption{Migration of Jupiter in a disc with viscosity $\alpha=10^{-4}$ for higher resolutions.} 
    \label{fig:res_a-4}
\end{figure}

\begin{figure}[h]
    \centering
    \includegraphics[width=\columnwidth]{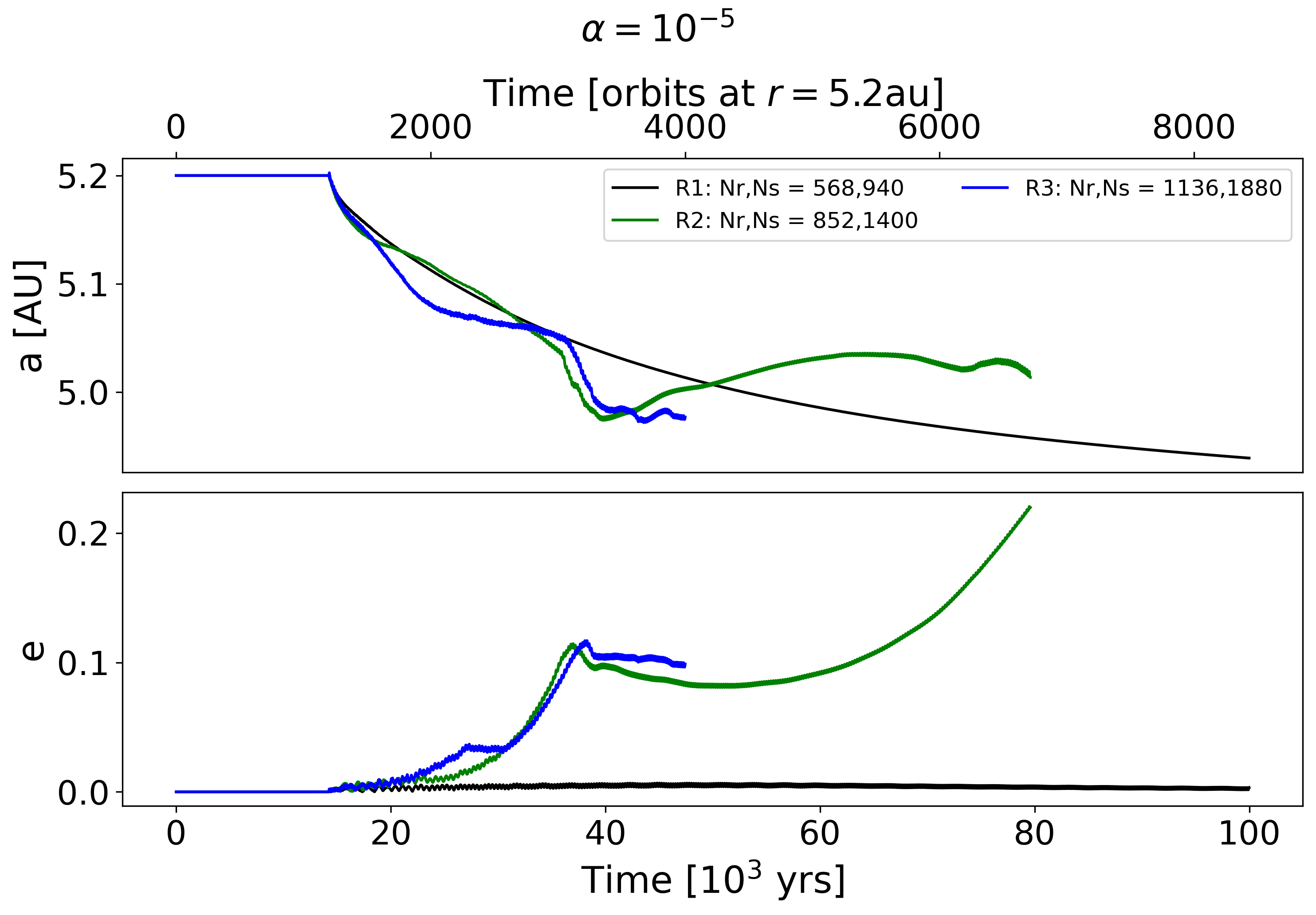}
    \caption{Migration of Jupiter in a disc with viscosity $\alpha=10^{-5}$ for higher resolutions.} 
    \label{fig:res_a-5}
\end{figure}

In this section, we first show the convergence of our results with higher resolution and secondly we show that, at lower viscosity values, such convergence cannot be achieved. We focus on the first phase of our simulations, that is when Jupiter is migrating alone in the disc. We recall that, in this paper, the resolution is such that in the nominal simulation we have about $7.4$ cells per scale height at $r_0$, corresponding to $N_{\rm{rad}} \times N_{\rm{sec}} = 568 \times 940$. We show in \cref{fig:res_a-4} that the simulation remains unchanged with the following grids:  $N_{\rm{rad}} \times N_{\rm{sec}} = 852 \times 1400$ and $N_{\rm{rad}} \times N_{\rm{sec}} = 1136 \times 1880$. \\
In the case of lower viscosity, however, we show in \cref{fig:res_a-5} that at higher resolution the migration of a single Jupiter-like planet is perturbed by small-scale instabilities in the discs. This confirms the findings from \cite{mcnally_migrating_2019}, claiming that numerical convergence cannot be obtained for values of viscosity of about $\nu < 10^{-7}$, corresponding in our {case h=0.035 to about $\alpha \simeq 3\cdot 10^{-5}$ (and $\alpha \simeq 4\cdot 10^{-5}$ in the case $h=0.05$)}.


\section{Removing gas potential: Effect on the eccentricities}\label{app:gasRem}

\begin{figure}[h]
    \centering
    \includegraphics[width=\columnwidth]{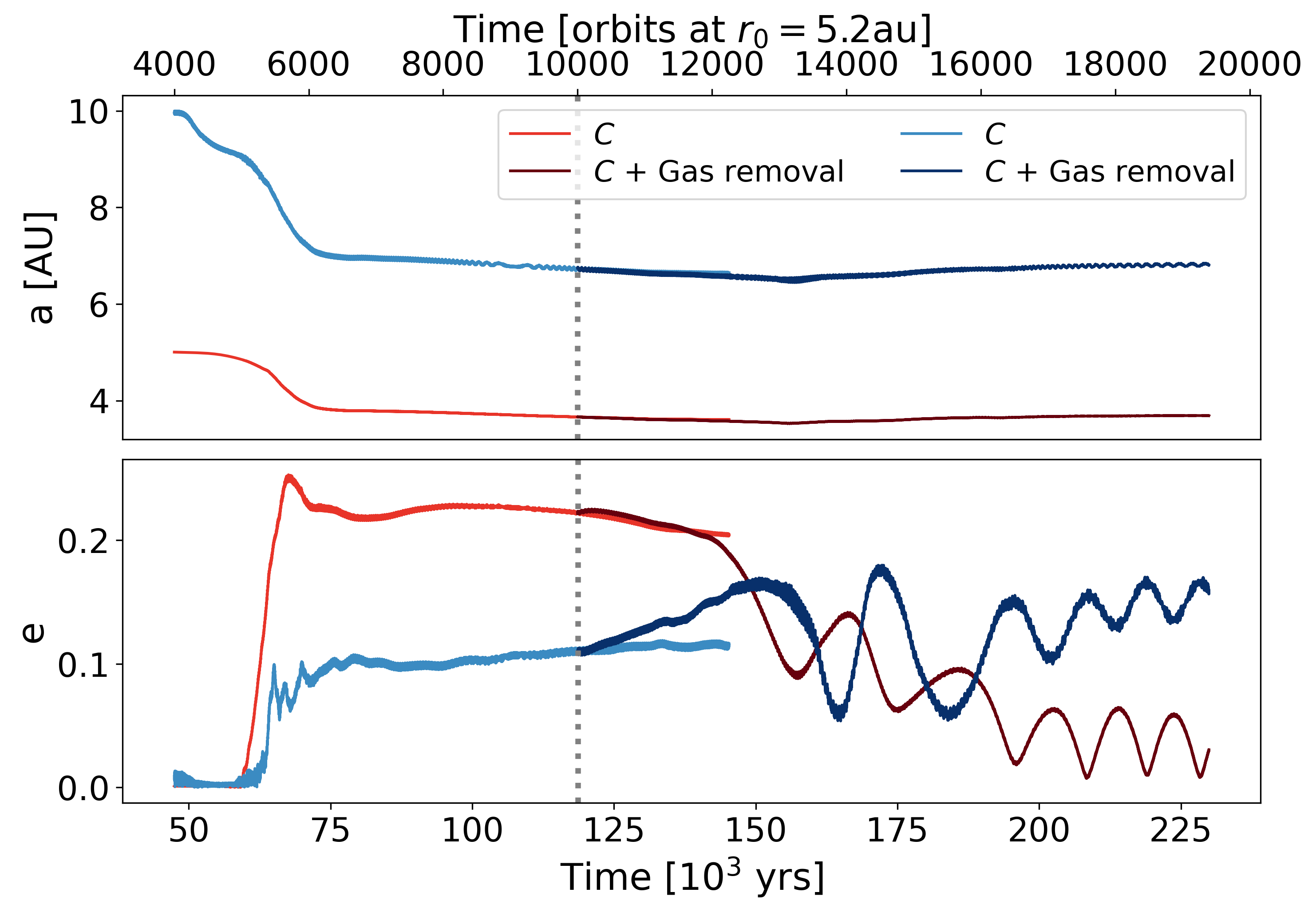}
    \caption{Orbital evolution of Jupiter and Saturn for simulation set $C$. The top and bottom panels show the semi-major axis and the eccentricity, respectively. The vertical dotted lines indicate the time $t_0$ at which we {started the simulation with gas removal, namely '$C$ + Gas removal'}. {The nominal simulation was continued for $t>t_0$ until a significant difference appeared with respect to the gas removal.}}
    \label{fig:gas_removal1}
\end{figure}

\begin{figure}[h]
    \centering
    \includegraphics[width=\columnwidth]{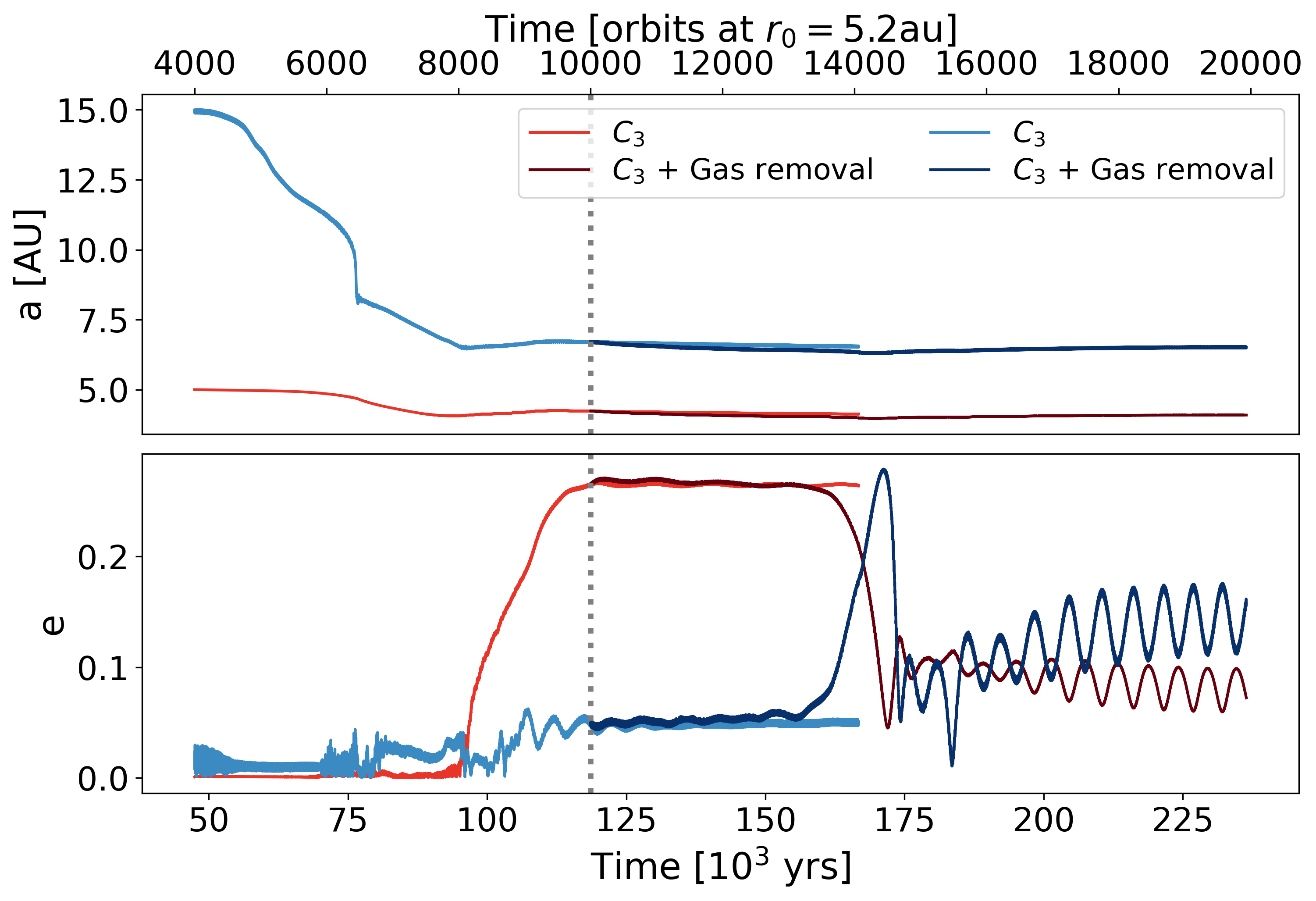}
    \caption{{Same as \cref{fig:gas_removal1}, but for simulation set $C_3$.}}
    \label{fig:gas_removal2}
\end{figure}

\begin{figure*}[h]
    \centering
    \includegraphics[width=\textwidth]{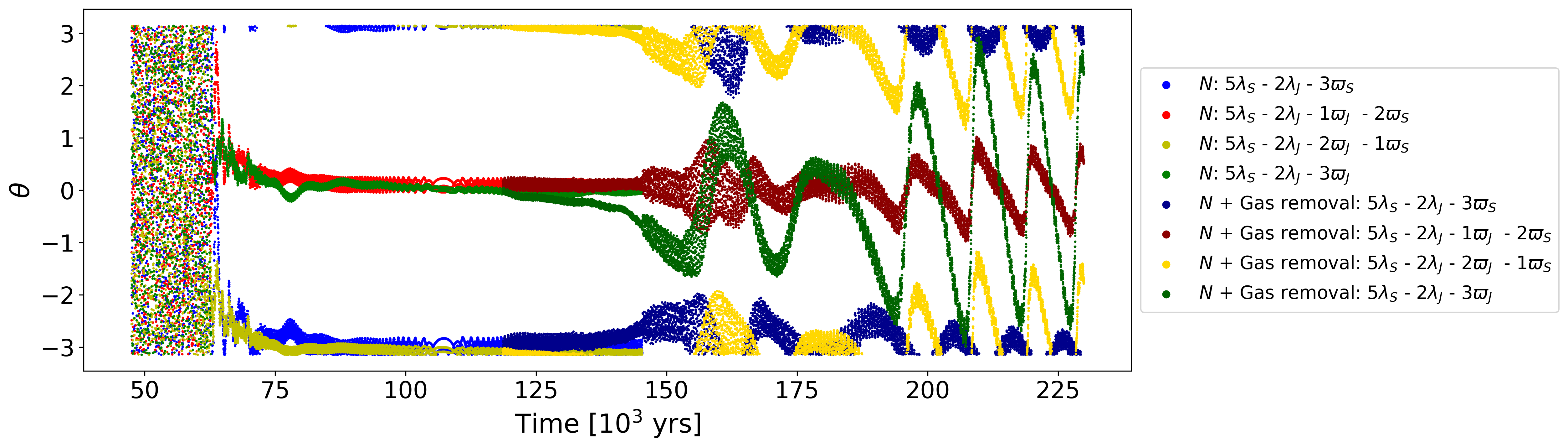} 
    
    \includegraphics[width=\textwidth]{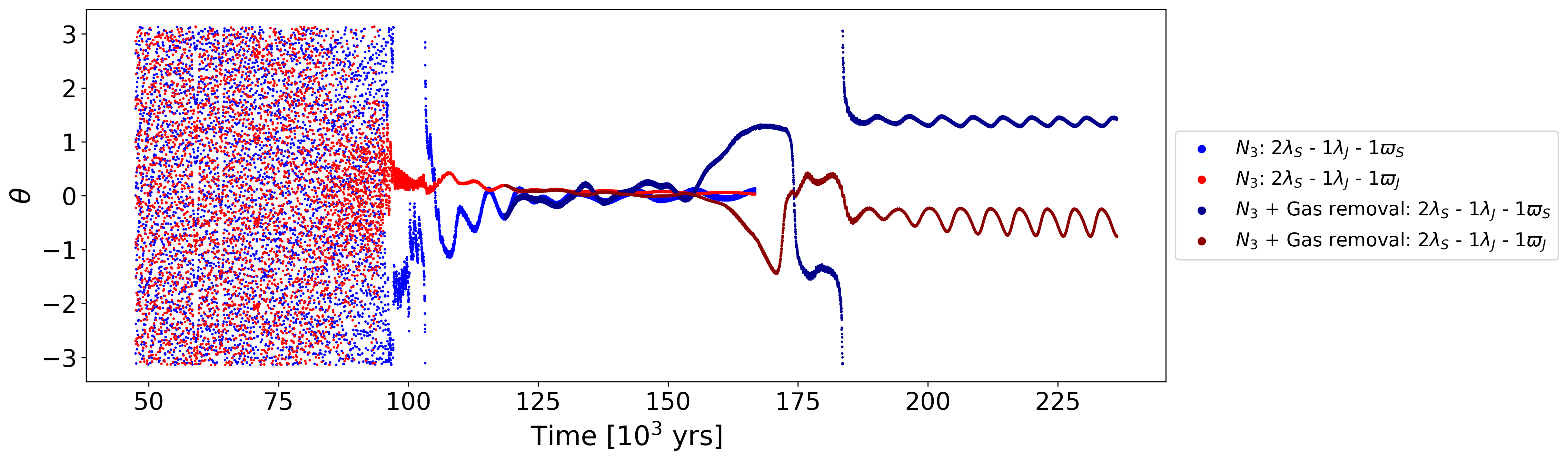}
    \caption{{Resonant angles corresponding to the simulations shown in \cref{fig:gas_removal1} in the top panel, and those corresponding to \cref{fig:gas_removal2} in the bottom panel. We find that in both cases of 2:1 or 5:2 MMR, the planets remain locked into resonance after the gas removal. Although, their resonant angles either librate with larger amplitude or no longer around 0.}}
    \label{fig:Resangle_Evap}
\end{figure*}

In many simulations, we have found that Jupiter's eccentricity is much higher than that of Saturn. We have suggested that these eccentricity values are maintained by the gas' potential. \\
We have tested this hypothesis by gradually removing  the force $\vec F(t)$ of the gas onto  the planets with a taper function 
\begin{equation}
\vec F(t)=\vec F(t_0)e^{-{\frac {(t-t_0)} {\tau}}}
,\end{equation}
with $\tau$ being the removal timescale and $t_0$ a suitable starting time. {We show} as an example in \cref{fig:gas_removal1} and \cref{fig:gas_removal2} the evolution of orbital parameters of simulation set $C$ and $C_3$.
At time $t_0 \sim 120000$ years, we restarted the
simulation removing the force  of the gas onto the planets over a timescale $\tau = 10^4$  orbits at $r_0 = 5.2$AU.  We observed that: \\
\begin{itemize}
    \item[i)] the pair of planets remains locked in their initial resonant configuration,
    \item[ii)] the eccentricity of Jupiter reduces and that of Saturn increases. This is consistent with the results obtained for $C_{{\Sigma -}}$ and more generally expected from N-body simulations,
    \item[iii)] the libration angles $\theta$ have a non-zero amplitude indicating that the system is no longer at the stable equilibrium point. {We note that the centre of libration can also change (e.g. in the case of the 2:1 MMR), as it is known to happen once the disc dissipates (see for example \cite{siegel_resonant_2021}).}
\end{itemize}
{Points i) and iii) are illustrated in \cref{fig:Resangle_Evap}.}

\section{Migration in the 3:2 MMR}\label{app:outward}

\begin{figure}[h]
    \centering
    \includegraphics[width=\columnwidth]{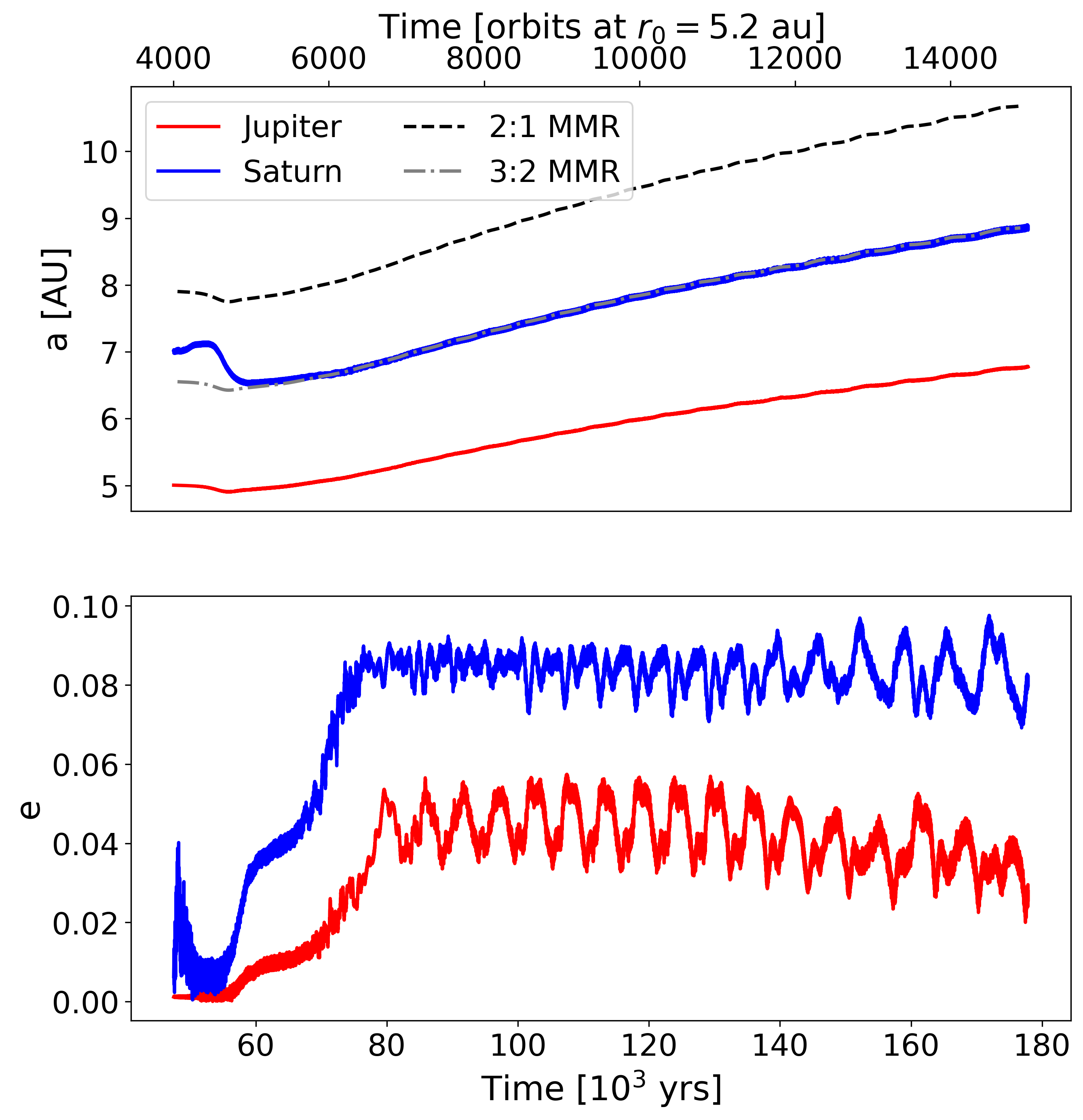}
    \caption{Migration of Jupiter and Saturn if Saturn was to form {on the inside of the of} 2:1 MMR. The top panel shows the evolution of the semi-major axes of Jupiter, in red, and Saturn, in blue. The dashed black and grey lines represent the position of the 2:1 and 3:2 MMR with Jupiter, respectively. The bottom panel shows the eccentricities of both planets.} 
    \label{fig:asaj14}
\end{figure}

{We have shown that a Saturn mass planet is not able to cross the 2:1 MMR. However we wonder what the outcome} of the migration of the pair of planets would be if Saturn were to form {on the inside of the} 2:1 MMR, so $a_S/a_J <1.58$. {In this case, we expect Jupiter and Saturn to be trapped in the 3:2 MMR, but the direction of their migration is an open question. Since the Masset and Snellgrove mechanism is independent of the viscous drift, we expect it to still be present in the low-viscosity case. However, in low-viscous discs, the gaps carved by planets are significantly deeper and it is not obvious that gas can flow from the outer to the inner disc, allowing the condition for outward migration (see \textit{(iii)} in \cref{sec:intro}) to be satisfied.}\\
Taking the same procedure as for the {previous} simulations, we placed Saturn at a distance of $a_S = 1.4 a_J$, {for $h=0.035$}. We show the result of this simulation in \cref{fig:asaj14}. The planet moved inwards until reaching the 3:2 MMR in which the pair was stably locked for the rest of the simulation. Once the planets were in this configuration, they migrated outwards together as was observed in classically viscous discs \citep{masset_reversing_2001, morbidelli_dynamics_2007-1}. We checked {that gas flows from the outward to the inner disc} by placing gas tracers at the edge of the common gap. We indeed see that even in the low-viscous case, gas flows from the outer to the inner disc allowing the planets to migrate outwards.\\
We therefore find that in low-viscosity discs, it is possible to find the Masset and Snellgrove mechanism resulting in the pair Jupiter-Saturn migrating outwards, but only when the pair is locked in the 3:2 MMR. To have such a configuration of the planets requires that Saturn forms at a distance less than $1.58 a_J$. This is not likely if Jupiter is already formed as that would require Saturn to form in the gap of Jupiter. Another possibility would be that both planets form at the same time, allowing them to grow close enough to be in the 3:2 resonance. However, we have not studied this case as this is out of the scope of this paper. 

\end{document}